\documentclass[12pt,preprint]{aastex}
\newcommand{\adeg}[1]{{#1}$^{\circ}$}
\newcommand{\amin}[1]{{#1}$^\prime$}
\newcommand{\asec}[1]{{#1}$^{\prime\prime}$}

\newcommand{\mjybeam}[1]{{#1}\,mJy\,beam$^{-1}$}

\newcommand{\fermi}{\emph{Fermi} }

\begin{document}

\title{Known Pulsars Identified in the GMRT 150 MHz All-Sky Survey}

\author{D. A. Frail\altaffilmark{1}}
\email{dfrail@nrao.edu}
\author{P. Jagannathan\altaffilmark{1,2}}
\author{K. P. Mooley\altaffilmark{3,4}}
\and
\author{H. T. Intema\altaffilmark{1,5}}

\altaffiltext{1}{National Radio Astronomy Observatory, 1003 Lopezville Road, Socorro, NM 87801, USA}
\altaffiltext{2}{Department of Astronomy, University of Cape Town, Private Bag X3, Rondebosch 7701, Republic of South Africa}
\altaffiltext{3}{Astrophysics, Department of Physics, University of Oxford, Keble Road, Oxford OX1 3RH, UK}
\altaffiltext{4}{Hintze Research Fellow}
\altaffiltext{5}{Leiden Observatory, Leiden University, Niels Bohrweg 2, NL-2333CA, Leiden, The Netherlands}

\begin{abstract}
We have used the 150 MHz radio continuum survey (TGSS ADR) from the Giant Metrewave Radio Telescope (GMRT) to search for phase-averaged emission toward all well-localized radio pulsars north of \adeg{$-$53} Declination. We detect emission toward 200 pulsars with high confidence ($\geq 5\sigma$) and another 88 pulsars at fainter levels.  We show that most of our identifications are likely from pulsars, except for a small number where the measured flux density is confused by an associated supernova or pulsar-wind nebula, or a globular cluster.  We investigate the radio properties of the 150 MHz sample and we find an unusually high number of gamma-ray binary millisecond pulsars with very steep spectral indices. We also note a discrepancy in the measured flux densities between GMRT and LOFAR pulsar samples, suggesting that the flux density scale for the LOFAR pulsar sample may be in error by approximately a factor two. We carry out a separate search of 30 well-localized gamma-ray, radio-quiet pulsars in an effort to detect a widening of the radio beam into the line-of-sight at lower frequencies. No steep spectrum emission was detected either toward individual pulsars or in a weighted stack of all 30 images. 
\end{abstract}


\section{Introduction}\label{sec:intro}

It is well known that pulsars have considerably steeper spectral indices than the background population of radio sources. Their flux density (S$_\nu$) can be described by a single power-law with slope $\alpha$ (i.e. S$_\nu\propto{\nu}^\alpha$).  Observationally-derived spectral indices have been determined variously to be in the range of $-1.6\pm{0.3}$ \citep{lylg95} to  $-1.8\pm{0.2}$ \citep{mkkw00}. \citet{blv13} attempted to remove pulsar survey biases to derive an intrinsic spectral index of $-1.4\pm{1.0}$. There have been claims that millisecond pulsars have more shallow spectral indices on average than ``normal" (i.e. non recycled)  pulsars \citep{kll+99}, but this may be due to an observational bias \citep{blv13}. 

Deviations from this pure power-law behavior have been seen at both high and low frequencies, with some fraction (10\%) having evidence for flat spectra and spectral steepening above several GHz \citep{mkkw00}. Low frequency turnovers first seen by \cite{sie73}, have now been measured for both normal and millisecond (MSP) pulsars, typically below 100 MHz \citep[e.g.][]{drt+13,kvl+15}. External free-free absorption, either in the immediate environment of the pulsar or along the line of sight, gives a good explanation for the origin of the high frequency turnover pulsars \citep{lrkm15,rla16}.  The origin of the low frequency turnovers, is not so clear. They could be telling us something fundamental about the energy distribution of the coherent emitting electrons, or the turnover could be due to absorption, occurring either within the pulsar magnetosphere or along the line of sight.

Progress in understanding these low frequency behaviors and their dependence (if any) on known pulsar parameters has been slow, owing to a shortage of flux density measurements below 1 GHz \citep{mgj+94}. The best efforts to date are those of \citet{mms00} who made measurements of 235 pulsars at 102.5 MHz, while 30 MSPs were observed at 102 and  110 MHz  \citep{kl01}.  Fortunately, the situation is changing with new instruments such as the LOw-Frequency ARray (LOFAR)  and the Long Wavelength Array (LWA). There are more recent LWA measurements of 44 pulsars from 10-88 MHz \citep{srb+15} and LOFAR has now observed large samples of normal and MSPs at 110-188 MHz and \citep{bkk+15, kvh+16}. 
 
In this paper we use the recently completed GMRT Sky Survey (TGSS ADR) to study known radio and gamma-ray pulsar populations at 150 MHz. This paper is arranged as follows. In \S\ref{survey} we briefly describe the TGSS ADR survey while in \S\ref{method} we outline our search methods for both the radio-loud and radio-quiet samples.  The results are discussed in \S\ref{results} where we derive estimates of the spectral index distribution of the TGSS ADR pulsars, and we compare the derived flux densities and the detection statistics with previously published samples. Our conclusions and suggestions for future work are given in \S\ref{sec:conclude}.

\section{The TGSS ADR Survey}\label{survey}

The Giant Metrewave Radio Telescope (GMRT) was used to carry out a radio continuum survey at a frequency of 150 MHz using a total of 2000 hrs of observations. The entire sky was surveyed in over 5000 partially overlapping pointings from $-55$\degr\, declination to the northern polar cap covering 37,000 deg${^2}$.

The entirety of these data have recently been re-processed \citep[TGSS ADR;][]{int16} creating high-quality images of approximately 90\% of the entire sky. The TGSS ADR achieves a median rms noise level of \mjybeam{3.5} and an angular resolution of \asec{25} for Dec.$>19^{\circ}$, and 
\asec{25}$\times$\asec{25}/cos(DEC-19$^\circ$) for more southern declinations. In the final catalog there are some 0.62 million radio sources down to the 7$\sigma$ level. Compared to existing meter-wavelength surveys \citep{lcv+14,hpo+15,wlb+15},  the TGSS ADR represents a significant improvement in terms of number of radio sources, sensitivity and angular resolution. The improved angular resolution in particular allows accurate matching of radio sources with counterparts at other wavelengths. The capabilities of the TGSS ADR are well-matched to existing surveys \citep{bwh95,ccg+98,bls+99} and provide a large frequency leverage arm for spectral index measurements. For more details on this survey and how to obtain the publicly available mosaic images and source catalog, see \citet{int16}.

The observing bandwidth and integration time of the survey are especially relevant to the detection of the phase-averaged emission from pulsars. The original data were recorded with 256 frequency channels across 16.7 MHz of total bandwidth centered on 147.5 MHz. 
% During final image processing these data were frequency-averaged from 64 channels of 0.26 MHz each to create a single Stokes-I broad-band channel. 
The GMRT visibility data from the archive were saved as 16.1 s averages. Typically each (pointing) direction on the sky was observed as a series of short snapshots 3 to 5 times over the course of a single night's observing. The total integration time was 15 minutes per pointing on average.
During imaging of the pointings, these data were combined in time and frequency to create a single Stokes-I image.
The full duration spent on any given point on the sky is more difficult to quantify. The final data products used for the analysis are 25 deg$^2$ image mosaics, formed by combining overlapping 7.6 deg$^2$ pointing images.
Note that some pointings were observed repeatedly during multiple observing sessions, imaged separately, and combined in creating the mosaics.
% Thus some sources in overlapping {areas} may be observed more than once, separated by days {or} even months. 
As a result, many sources have been observed more than once, sometimes separated by days or even months.
As the survey's pointing centers followed the FIRST survey hexagonal grid strategy \citep{bwh95}, the TGSS ADR will have similar duration statistics \citep[see][]{thw+11}.

\section{Methods\label{method}}

\subsection{Radio Loud Pulsars}\label{sec:loud}

While pulsars are typically detected by their pulsed, periodic emissions, they can also be identified in interferometric images as phase-averaged continuum point sources. \citet{kcac98} were the first to employ a wide-field radio survey for this purpose. They used the 1.4 GHz NRAO VLA Sky Survey \citep[NVSS;][]{ccg+98} to identify 79 known pulsars from the total intensity alone, while \citet{ht99} added in the polarized intensity to identify 97 pulsars from the same survey. The 325 MHz Westerbork Northern Sky Survey  \citep[WENSS;][]{rtb+97} was used by \citet{kou00} to find radio emission toward 25 known pulsars. For this project we employ the TGSS ADR at 150 MHz, using a version of the source catalog that was formed by running the source extraction algorithm PyBDSM \citep{mr15}\footnote{http://www.astron.nl/citt/pybdsm/} with its default parameters searching the mosaicked images down to a 5$\sigma$ detection threshold. For our list of known pulsars we used the HEASARC 27 December 2015 version of the ATNF Pulsar Catalog \citep{mhth05}. A total of 1238 pulsars were selected with dec.$\geq-52^\circ$ and having known positions with $\Delta$dec.$<\pm{3^{\prime\prime}}$ or $\Delta$R.A.$<0.35^s$. {A history file listing the contents and any changes to the ATNF database as of December 2015 is at their web site\footnote{http://www.atnf.csiro.au/research/pulsar/psrcat/catalogueHistory.html}. Up to approximately November 2015, our sample included all well-localized normal pulsars from the \fermi sample at Stanford University as well as the MSP sample at the University of West Virginia.} The pulsar positions were corrected for proper motion using the mean epoch of the TGSS ADR of 11 January 2011 (MJD\,55579.0). 

Following \citet{hwb15}, we searched for matches between the PSR and TGSS ADR catalogs out to a radius of 30\% of the FWHM of the 25$^{\prime\prime}$ beam, or 7.5$^{\prime\prime}$. We find 200 known pulsars, or 16\% of the sample, are associated with TGSS ADR sources.  We list these detections in Table \ref{tab:bright} along with some basic pulsar parameters (period and dispersion measure) along with the total flux density and peak flux from the TGSS ADR.  {The ratio of the total flux density and the peak flux can be a useful proxy in helping decide whether the radio emission is extended or unresolved, and therefore likely phase-averaged pulsar emission. We return to this point, as well as discussing the rates of false positives in \S\ref{ids}.} For all matches, we derived two-point spectral indices between the TGSS ADR total flux densities at 150 MHz and the 1400 MHz values from the pulsar catalog. The 400 MHz flux density was used in those few cases where the 1400 MHz values were missing. When no values were provided in the pulsar catalog database, we obtained flux densities from the original literature.

\begin{figure}
\plotone{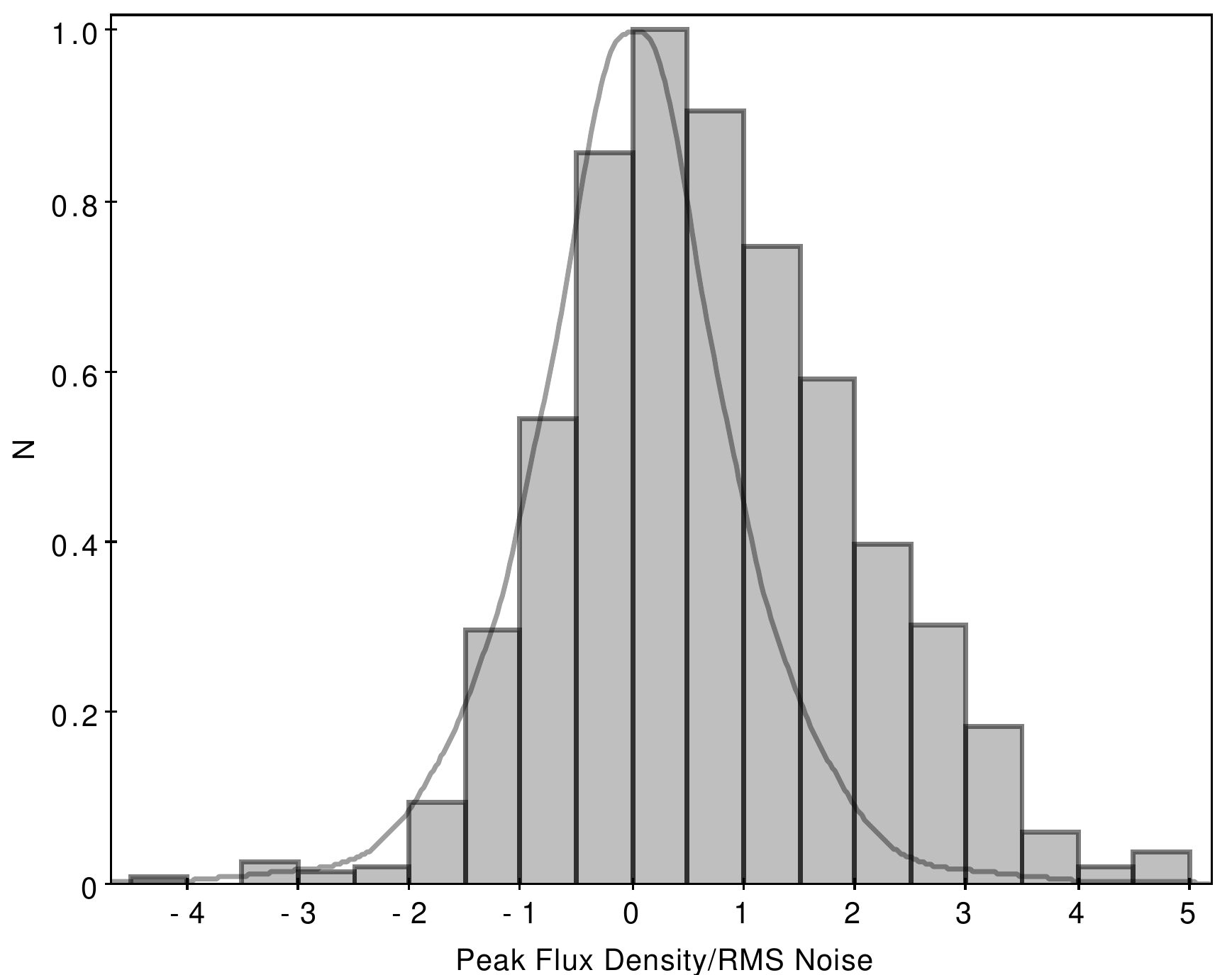}
\caption{The distribution of the signal-to-noise toward the pulsars that are not in the bright TGSS ADR 5$\sigma$ sample. The peak flux
has been measured at the position of each pulsar and a local estimate of the rms noise has been determined. The bins are specified in units of $0.5\sigma$. The distribution is normalized to the peak. The thin line is an estimate of the blank sky contribution estimated by making peak and rms noise estimates at several locations within a radius of a few arcminutes around each pulsar.}\label{fig:noise}
\end{figure}

The search method above was supplemented with an image-based approach for pulsars below the 5$\sigma$ limit of the catalog. For all  of the original 1238 well-localized pulsars, we measured the peak flux in the TGSS ADR images at the pixel corresponding to the pulsar position, along with an estimate of the rms noise in immediate vicinity. We did not attempt to search over some radius and fit a Gaussian since below 5$\sigma$ such a approach would likely lead to many more false positives. The positive identifications are defined as having S$_p$/$\sigma_{rms}\geq$2.5. Our justification for this choice of threshold is shown in  Fig. \ref{fig:noise}. We make an estimate of the shape of the signal-to-noise distribution as an estimate of the blank sky in the vicinity of these pulsars (solid line). The positive S/N peaks are strongly skewed above that expected from Gaussian noise. From the ratio of the levels of the positive and negative 2.5$\sigma$ bins we estimate that 4\% of the detections at this level will be false positives, or about 2 sources.

With this image-based approach we find significant emission towards another 88 pulsars. For each of these we inspected the mosaic images to verify that the emission was coming from a point centered on the pulsar position and was not due to a nearby extended source or an image artifact. Table \ref{tab:faint} lists the peak flux and rms toward all 88 pulsars, along with some basic pulsar parameters identical to those in Table \ref{tab:bright}.  Given the lower significance, we are not as confident in these identifications as we are with those in Table \ref{tab:bright}.

%The skewed S/N distribution in Fig. \ref{fig:noise} relative to the background suggests that there is significant signal in the direction of the remaining pulsars. To demonstrate this we employed a stacking analysis of all the sources below our detection limit \citep{whb+07}. We made a weighted sum of TGSS ADR image cutouts of all pulsars below the 2.5$\sigma$ cutoff. The final stacked image is shown in Fig. \ref{fig:faint}. There is a statistical detection of ...GIVE DETAILS. As noted by \citet{msc+12} this method is biased toward sources just below the detection limit but it is useful enough to show that the skewed distribution in Fig. \ref{fig:noise} is produced by real sources.

\subsection{Radio Quiet Pulsars}\label{quiet}

Motivated by recent claims of the detection of pulsed radio emission from the ``radio-quiet'' PSR\,J1732$-$3131 \citep{mad12}, we carried out a search for emission at 150 MHz. In the Second \fermi Large Area Telescope (LAT) Pulsar Catalog \citep[2PC;][]{aaa+13} there are 35 PSRs that are radio quiet, defined as having a phase-averaged flux density S(1.4 GHz)$\leq 30$ $\mu$Jy. In the meantime the sample has grown; an updated list of radio quiet pulsars is available on the LAT team web site\footnote{\url{https://confluence.slac.stanford.edu/display/GLAMCOG/Public+List+of+LAT-Detected+Gamma-Ray+Pulsars}}. For accurate pulsar positions we began with the recent compilation of \citet{krj+15}, which has a table of positions for normal pulsars and MSPs obtained from both timing and multi-wavelength observations. All positions from \citet{krj+15} are computed at the epoch MJD 55555 (25 Dec 2010). This is useful when comparing to the TGSS ADR which was observed around the same epoch. Additional X-ray positions are taken from \citet{mmd+15}. 

The final list consisted of 30 radio quiet pulsars in the declination range of TGSS ADR with localizations of an arcsecond or better (Table \ref{tab:unknown}). For all pulsars we extracted image cutouts and looked for faint point sources at the pulsar position. We then made a final stacked image of all 30 pointings weighted by the inverse square of the local rms noise for each image. As the image pixel size is the same \asec{6.2} in all images, this ensures accurate image stacking.  No source is detected. The rms noise is \mjybeam{0.7} and the max/min on the image is  approximately $\pm$\mjybeam{2.3}. 

\section{Results and Discussion\label{results}}

\subsection{Radio Loud Pulsars}

\subsubsection{Identifications}\label{ids}

The majority of the emission that was detected at 150 MHz is likely due to phase-averaged pulsar emission. In support of this we note that the distribution of PSR-TGSS ADR offsets follows the expected Rayleigh distribution, with 95\% of the identifications matched within a \asec{4.5} radius. This is consistent with the astrometric accuracy (68\% confidence) derived for the TGSS ADR of \asec{1.55} \citep{int16}. Given the source density of the TGSS ADR at the completeness limit of 17.6 source/deg$^2$, and a search of 1238 positions each of radius \asec{7.5} 
(see \S\ref{sec:loud}), we expect less than one false positive (i.e. a background radio source not associated with a pulsar).

\begin{figure}
\plotone{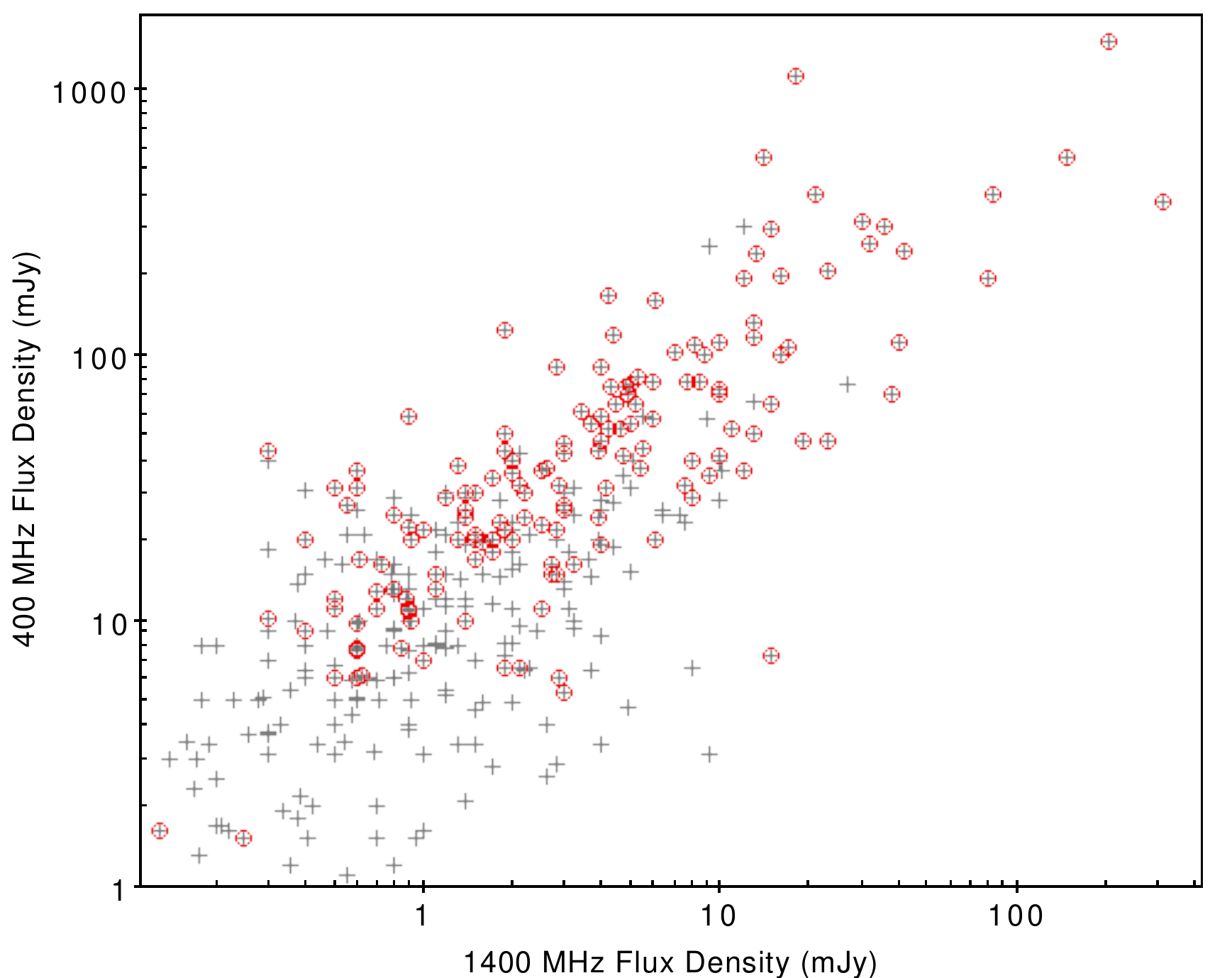}
% MADE IN TOMCAT WITH two files. flux-fluxFig.dat and psrcat-allDec15-prime.fits. 
\caption{The phase-averaged flux densities of known pulsars at 1.4 GHz and 400 MHz (crosses) as taken from the ATNF Pulsar Catalog. Circles indicate those pulsars towards radio emission was detected above $5\sigma$ in the TGSS ADR catalog at 150 MHz.}\label{fig:ff}
\end{figure}

Further support for pulsar identifications comes from Fig.\ref{fig:ff}, in which we the show all pulsars (crosses) with published 400 MHz and 1400 MHz flux densities in the ATNF catalog. Those pulsars with TGSS ADR detections are indicated by circles. This figure shows that the TGSS ADR associations are well-correlated with the brightest pulsars, and thus the number of false positives are likely to be low. Furthermore, the number of associations drop off sharply for S$_{1400}<0.6$ mJy and S$_{400}<5$ mJy, as would be expected for steep spectrum pulsars given that the median noise of the TGSS ADR is \mjybeam{3.5}. The two outliers in the bottom left corner of Fig.\ref{fig:ff} are PSR\, J2229+6114 and the LOFAR-identified PSR\, J0613+3731. Their flux densities at 150 MHz appear to be dominated a pulsar wind nebula called `The Boomerang" \citep{kru06} in the first case and some unidentified extended emission in the second case. 

As the above example illustrates, not all the matches in the TGSS ADR are from phase-averaged pulsar emission. Some radio emission is due an associated nebula (e.g. Crab), or is an from an ensemble of pulsars in a globular cluster.  {In \citet{int16} we derive an empirical formula to help decide when a radio source is unresolved. In the high signal--to-noise case this reduces to S$_t/$S$_p\leq 1.13$. However, some caution is needed in applying this criteria to pulsars since they can show strong time-variability during an integration time, violating one of the central assumptions of the van Cittert-Zernike theorem upon which radio interferometric imaging is based. This can lead to deviations in the Gaussian fitted beam, or in especially strong cases, diffraction spikes around the pulsar. A visual inspection of the images is required to be sure since this same condition is likely met by strongly scintillating pulsars like PSR\,B1937+21. We examined the images of {\it all} of the detections in Table \ref{tab:bright} and find that likely non-pulsar candidates are those entries for which the total flux density exceeds the peak flux (i.e. S$_t>$S$_p$) by more than 50\%.}  For those small number of TGSS ADR detections (11) that we suspect are contaminated in this way, we add a comment in Table \ref{tab:bright} and we do not derive a spectral index.

\subsubsection{Spectral Index Distribution}

\begin{figure}
\plotone{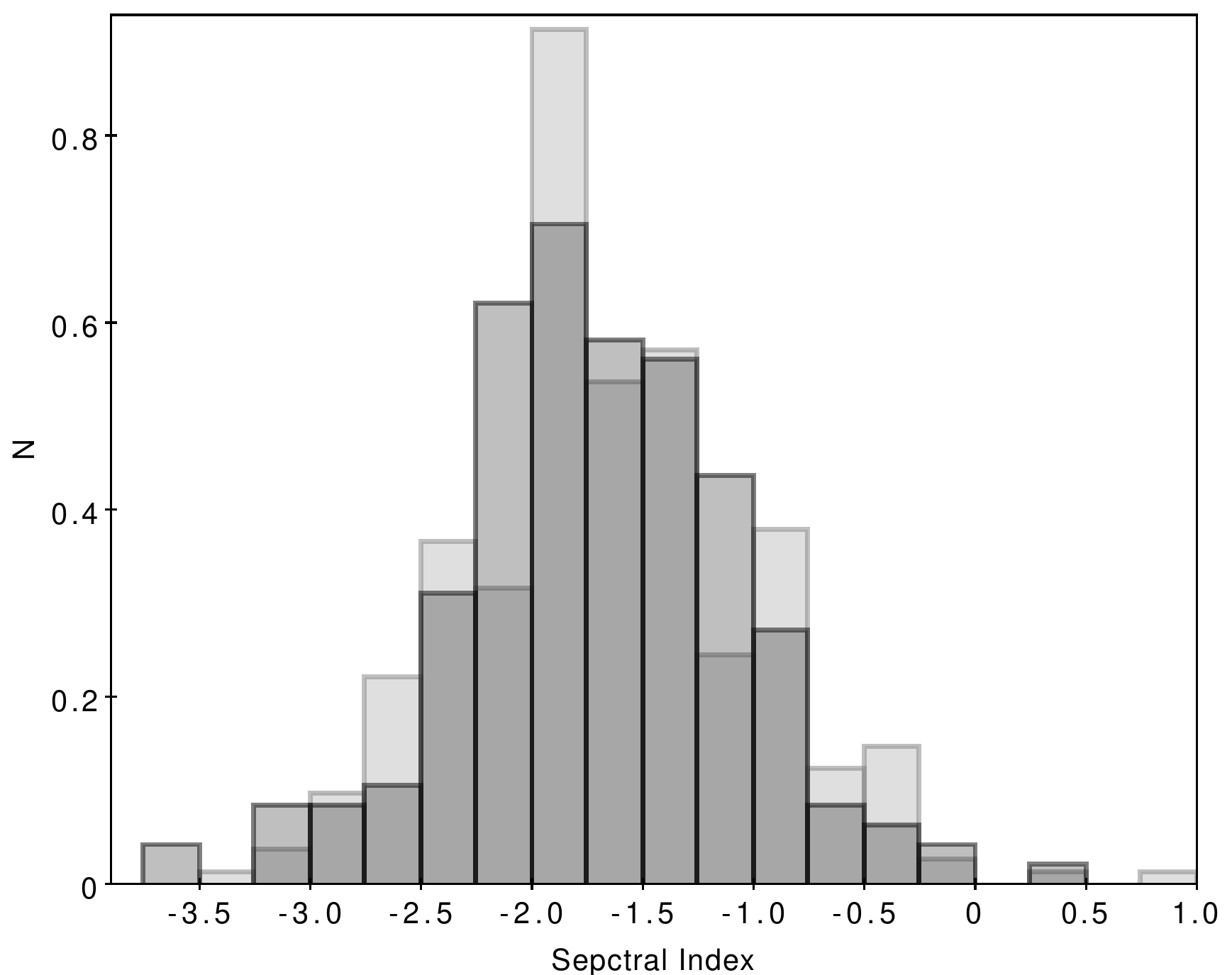}
\caption{The spectral index distribution of the ATNF Pulsar Catalog (light grey) versus the TGSS ADR $5\sigma$ sample (dark grey). Both samples have been normalized by total area for easier comparison. }\label{fig:spec_histo}
\end{figure}

The distribution of the two-point spectral indices of the TGSS ADR sample from Table \ref{tab:bright}  is shown in Fig. \ref{fig:spec_histo}. For comparison we have plotted the more comprehensive sample of 329 pulsars from the ATNF Pulsar catalog with non-zero spectral indices. As expected, the two histograms are in reasonable agreement with each other, both in terms of the width and the median of the two distributions. 

\begin{figure}
\plotone{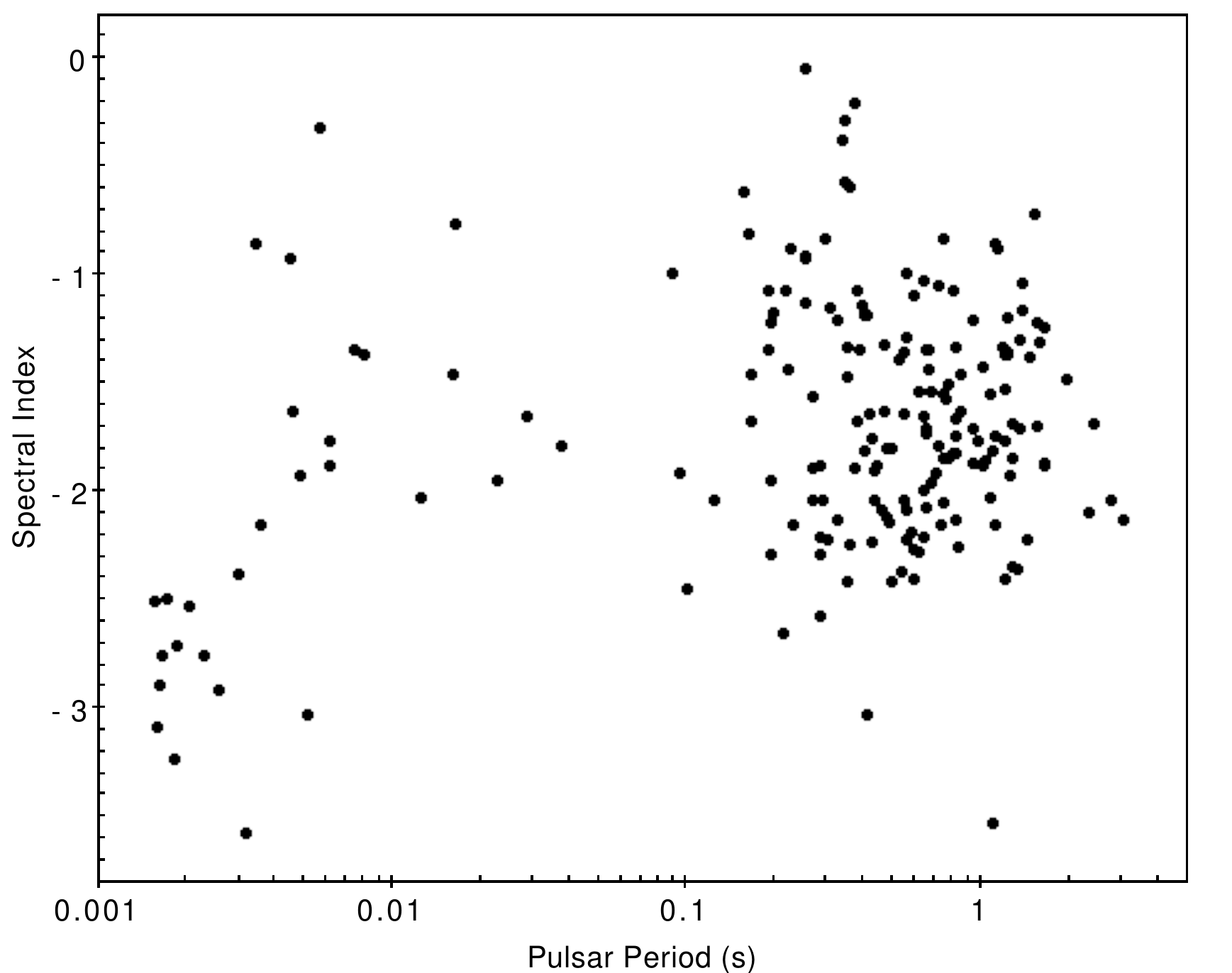}
\caption{The spectral index distribution of the TGSS ADR $5\sigma$ pulsars as a function of their rotational periods.}\label{fig:alpha}
\end{figure}

If we order the spectral index values by pulsar period (Fig. \ref{fig:alpha}) an unusual feature of our 150 MHz sample appears. The steep spectrum tail of the $\alpha$ distribution measured at low frequencies is dominated by short period pulsars. This effect is not seen in the ATNF pulsar catalog.  We have detected many of the fastest rotating MSPs at 150 MHz, and these pulsars show a marked preference for steeper spectral index values. Of the 16 pulsars with $\alpha<-2.5$, all but four are MSPs. Of these MSPs, all except one has been detected by the \fermi gamma-rays mission including several eclipsing MSPs such as PSR\,J1816+4510, with the steepest spectral index in Fig. \ref{fig:alpha}. The 18 MSPs in Table \ref{tab:faint} do not have ultra-steep spectra. \citet{kvl+15} were the first to note a tendency for the gamma-ray MSPs to be steep-spectrum outliers based on a smaller sample. 

Since the values in Table \ref{tab:bright} and Fig. \ref{fig:alpha} are two-point values, we suspected measurement error as the source of these large values. As a first step we re-calculated the spectral index of all pulsars with $\alpha<-2.5$ using the flux density and observing frequency taken from the original references. If no rms noise was given we assumed a fractional error of 50\% for the flux density when estimating the uncertainty on $\alpha$.

There are several useful compilations of flux density measurements and spectral indices we can use to cross check our measurements \citep{tbm+98,kxl+98,kvl+15}. We find reasonable agreement in the $\alpha$ values for all of the MSPs within the errors. For PSR\,J1816+4510, the pulsar with the steepest 2-point spectral index in our sample, we re-fit our 150 MHz measurement along with a value at 74 MHz \citep{kvl+15} and flux densities at 350 and 820 MHz \citep{slr+14}. The latter two measurements were estimated from the radiometer equation so we have taken typical errors of $\pm$50\% on these two values. The mean spectral index is $-3.46\pm0.10$ in agreement with a preliminary value from \citet{kvl+15}.

Since there is no evidence that the distribution of MSP spectral indices is steeper than the general population \citep{tbm+98,kll+99}, we suspect this trend is the result of some low frequency bias.  Most of the steep spectrum MSPs in Table \ref{tab:bright}, were discovered in low frequency searches \citep[e.g.][]{fst+88,bhl+94,hrm+11,slr+14}. Two pulsars (B1937+21 and J0218+4232) had such steep spectral indices that they were initially identified in imaging data \cite[e.g.][]{nbf+95}. Thus it is reasonable to expect that the TGSS ADR survey at 150 MHz would be sensitive to steep-spectrum radio sources, with a similar bias as low-frequency searches for pulsations \citep{blv13}. This explanation, however, does not account for the preponderance of gamma-ray pulsars among our sample, nor for the unusually large fraction of (eclipsing) binaries. We know of no intrinsic property of the MSP population that would produce such an effect. {\citet{ckr+15} noted that the nearby MSPs were susceptible to deep flux density variations at decimeter wavelengths, with strong expoential statistics such that the measured median flux density is less than the mean, skewing the spectral index to steeper values.} Since many of these systems have been found within the error ellipses of \fermi unassociated sources \citep[e.g.][]{krj+15}, a more prosaic explanation may be that the \fermi mission has been such a prolific source of MSPs that they are over-represented in any sample. 

\subsubsection{Comparison with the LOFAR Sample}

It is illustrative to compare the TGSS ADR and LOFAR samples. While both surveys were undertaken at the same frequency, they were observed in very different ways.  Thus a comparison could give us some insight into the different biases of each survey. LOFAR has carried out a search for pulsed emission from all northern radio pulsars \citep{bkk+15,kvh+16}. This census was primarily  conducted with the LOFAR high-band antennas (HBA) between 110 and 188 MHz, with 400 channels each of 0.195 MHz in width, or a bandwidth of 78 MHz. Each pulsar was observed once for at least 20 minutes, although long period (P$>3$ s) normal pulsars and faint MSPs were observed up to 60 minutes in duration. Pulsed emission from a  total of 158 normal pulsars and 48 MSPs were detected. The GMRT observing method is summarized in \S\ref{survey} and the pulsar yield is given in \S\ref{ids}. We find 92 pulsars commonly detected in both the  LOFAR and TGSS ADR surveys (Tables \ref{tab:bright} and \ref{tab:faint}).

\begin{figure}
%\plotwo{tgss-lofar.pdf}
\plottwo{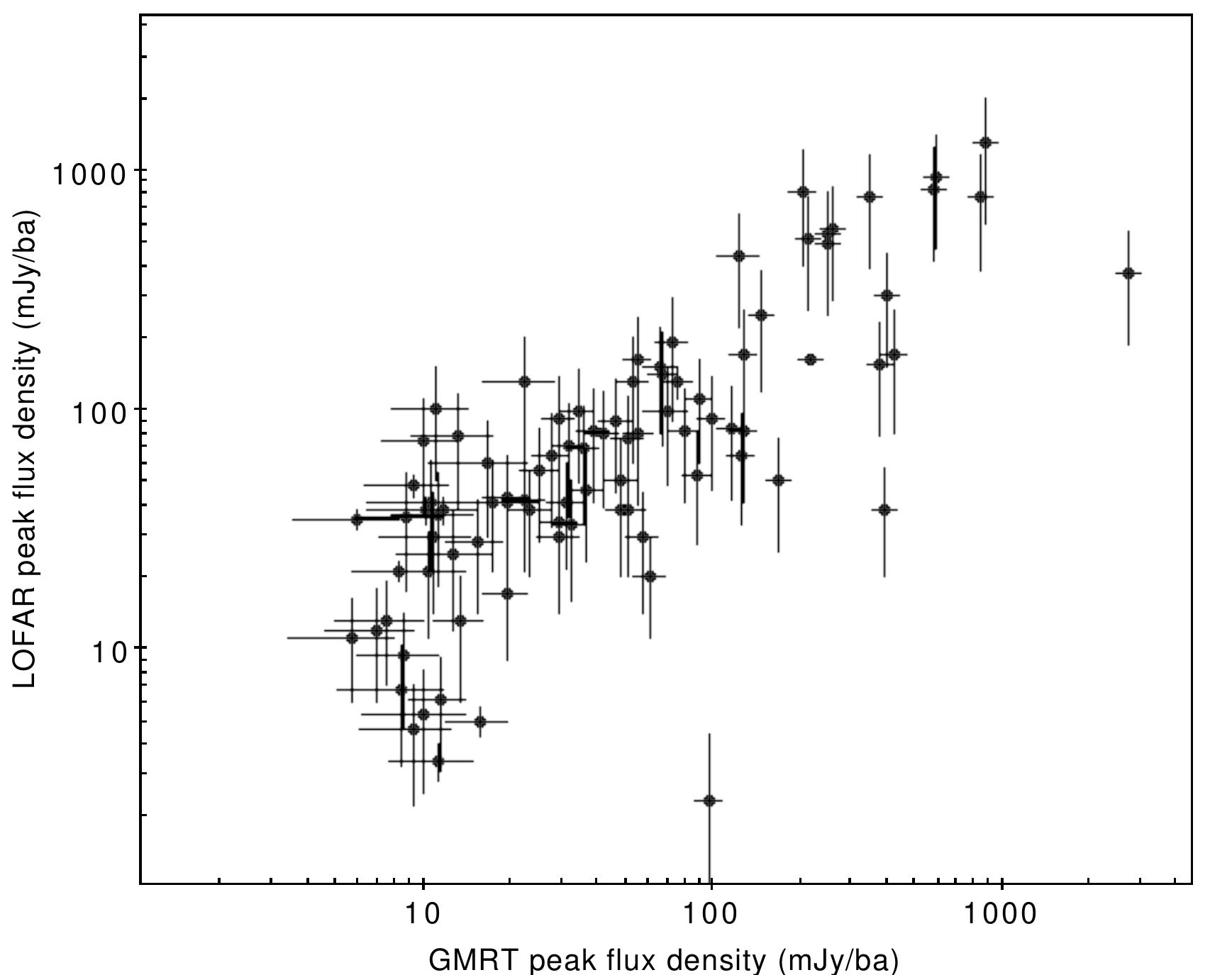}{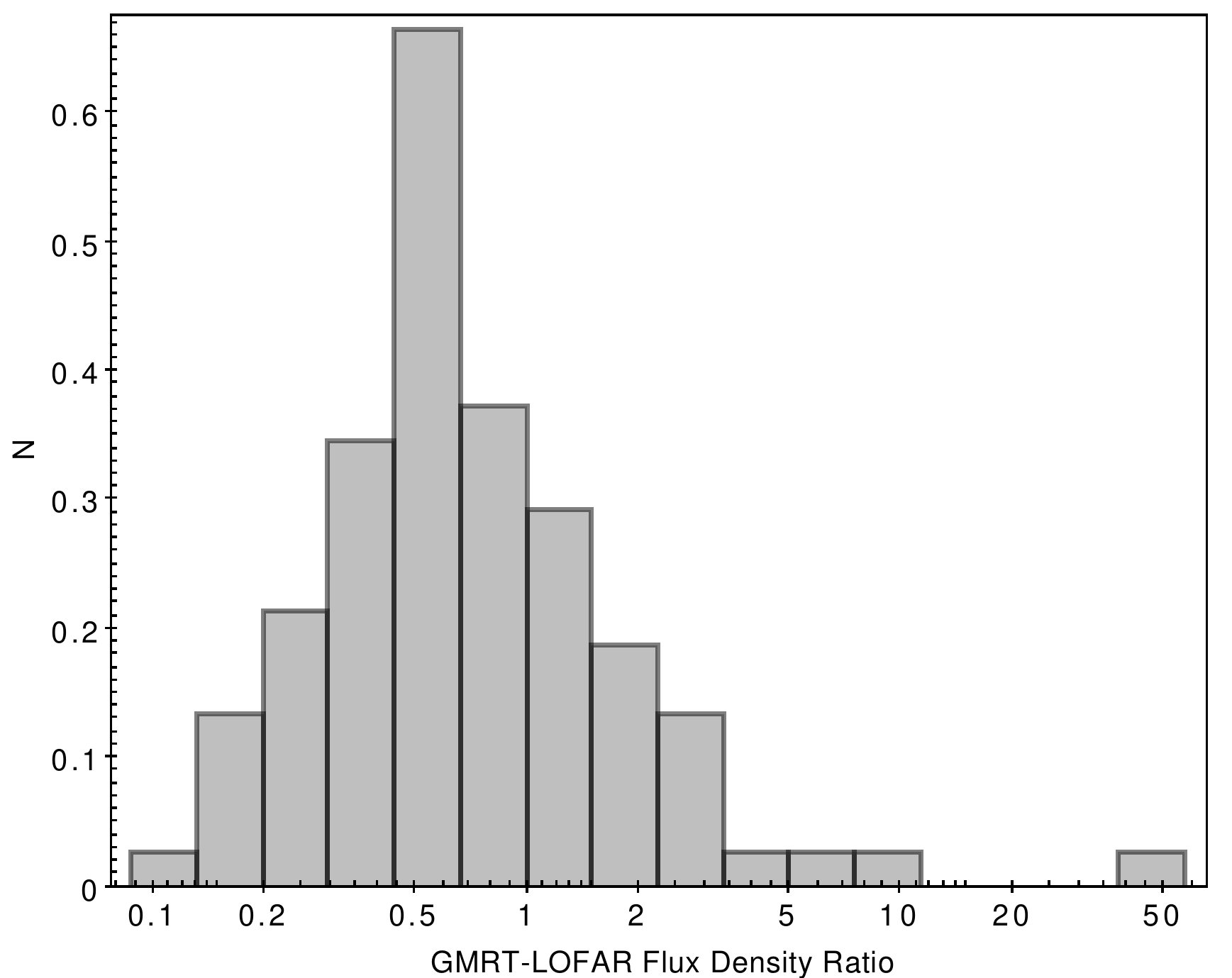}
\caption{(Left) The flux densities of the pulsars detected in common between the GMRT and LOFAR samples. A 10\% error has been added in quadrature to the measured GMRT errors quoted in Tables \ref{tab:bright} and \ref{tab:faint} as a conservative estimate of the systematic error in the flux density scale. The LOFAR errors  are taken from the original papers. (Right) The same data but plotted as a histogram of the ratio of the GMRT pulsar flux densities over the equivalent LOFAR values. The area under the histogram has been normalized by the total number of pulsars.}
\label{fig:compare}
\end{figure}

Figure \ref{fig:compare} (left) is a flux-flux plot of LOFAR and TGSS measured flux densities, while the same figure (right) shows a flux ratio plot of the same sample. The flux densities of the LOFAR and TGSS ADR pulsars do not agree. On average, the LOFAR pulsars are about two times brighter than the the TGSS ADR. The result persists even if we use only the bright pulsars in common (i.e. Table \ref{tab:bright}). There are some significant outliers, dominated by bright, scintillating MSPs such as PSR\,B1937+21 and PSR\,J0218+4232, but the overall trend is clear. 

We can immediately rule out frequency-dependent effects for this difference in the flux density scales since the surveys were performed at similar frequencies. Spectral curvature was the most likely explanation offered by \citet{kvh+16} for why the LOFAR flux densities for one third of their MSPs are {\it lower} than the predicted values based on an extrapolation from higher frequencies. Diffractive scintillation, while clearly important for the outliers, is not the likely origin for the systematic difference. The large observing bandwidths and the long integration times relative to the scintillation values for both the LOFAR and GMRT observations (\S\ref{survey}) suggest modulation of the flux density is not widespread; see \S\ref{sec:missing} and Appendix A of \citet{bkk+15}. There is one important difference: the typical 20-min LOFAR integration time is a single integration, while the 15-min GMRT observations are typically subdivided into 3--5 short observations taken over a night of observing. The later is a more optimal detection strategy when there are intensity variations caused by the phase fluctuations in the interstellar medium \citep{cl91}. If this effect is important, however, it would result in the LOFAR flux densities being {\it lower} on average that the GMRT values, the opposite of what is seen. Temporal scattering can also reduce the measured flux density for pulsed surveys but as an imaging survey, the TGSS ADR is not sensitive to pulse smearing caused by interstellar scattering. While the LOFAR surveys are sensitive to such effects, they would also act to {\it lower} the measured flux density. 

We are left with instrumental effects associated with gain calibration. The TGSS ADR flux density scale is good to about 10\% over the full sky.
%, and it is likely better. 
Taken in interferometric imaging mode, the data each day were calibrated back to several low frequency primary flux density calibrators (3C\,48, 3C\,147, 3C\,286 and 3C\,468.1). After calibration of the full survey, the accuracy of the flux density scale was cross-checked against other sky surveys such as 7C \citep{hrw+07} and the LOFAR Multi-frequency Snapshot Survey \citep[MSSS;][]{hpo+15} and they were found to agree at the $\sim$5\% level. On the other hand, the flux density calibration for the LOFAR pulsar survey was done directly using the radiometer equation for direction-dependent estimates of the antenna gain and the sky system temperature. The calibration was cross-checked with regular observations of a sample of normal pulsars and MSPs with well-determined spectra. Variations at a level 2--4 times larger than expected from scintillation alone were seen to occur and thus the resulting flux density scale was quoted with errors of $\pm$50\%. 

We tentatively suggest that our TGSS ADR pulsar sample shows that there remains an unaccounted gain error in the LOFAR pulsar observing system that results in an overestimate of the flux density scale by about a factor of two.

\subsubsection{The Missing Pulsars}\label{sec:missing}

Despite the high yield, there are also a number pulsars in Fig. \ref{fig:ff}  with large {decimeter} flux densities but with no TGSS ADR counterpart in Table \ref{tab:bright}. Likewise, we failed to detect several bright pulsars which had been found in previous low frequency pulsation surveys \citep[e.g.][]{kl01,bkk+15}. To investigate the origin of these missing pulsars, we defined a radio-bright sample from the original 1238 well-localized pulsars in \S\ref{sec:loud} as having 400 and 1400 MHz flux densities greater than 21 mJy and 1.8 mJy, respectively.  For a canonical pulsar spectral index these flux densities extrapolate at 150 MHz to the completeness limit of the TGSS ADR \citep{int16}.  There are 232 such pulsars. Of this sample, 70\% are detected and are listed in Tables \ref{tab:bright} and \ref{tab:faint}.

We can identify three possible reasons that about one third of this radio-bright sample of pulsars would not be detected in the TGSS ADR. The local rms noise may be too high, the pulsar spectrum may be flat or turn over at 150 MHz, or the signal may be reduced due to interstellar scintillation. It may be possible that one of these effects dominant or they are working in tandem. We will look at each of these in turn.

At low radio frequencies the synchrotron and thermal emission from the Galactic plane makes a non-negligible contribution to the system temperature of the receivers. The frequency dependence of the brightness temperature goes approximately at T$_b\propto\nu^{-2.6}$ \citep{hks+81} so unless the pulsar spectrum is steeper than this value, they become increasingly more difficult to detect. While the increased brightness temperature affects pulsed and imaging searches equally, the later also suffers from increased rms due to confusion and reduced image fidelity in the presence of bright Galactic HII regions or supernova remnants.  The pulsar B\,2319+60 is a good example of a bright pulsar confused by nearby bright, extended emission. We looked at the rms noise statistics of the  detected and non-detected samples, following up the large rms cases with a visual inspection of the TGSS ADR image data at the PSR positions. We find evidence that the rms noise of the images has some influence on the detectability of the pulsars. The median rms noise for the detections is \mjybeam{3.5}. while for the non-detections it is nearby twice this value (\mjybeam{6.9}).

The intrinsic spectral shape of the pulsar emission will also affect the detectability at low frequencies. The mean pulsar spectral index, while steep, has a wide scatter (\S\ref{sec:intro}). Likewise, for approximately 10\% of known pulsars there is evidence of a low-frequency spectral turnover, typically around 100 MHz (\S\ref{sec:intro}). Our 150 MHz sample has a number of pulsars with known spectral turnovers including PSR\,J2145-0750 \citep{drt+13}. We lack a large public database of accurate pulsar flux densities that would be sufficient to look for a turnover frequency for our non-detections, but fortunately most of them have single power-law measurements in the ATNF pulsar catalog. The median spectral index for the detections is $\alpha=-1.9$ and there are no pulsars in this sample as shallow as $\alpha\leq-0.5$. The non-detections have a much flatter median spectral index of $\alpha=-1.3$. At least one third of our non-detections have spectral indices that are so flat that we do not expect to detect them at 150 MHz based on an extrapolation of their 400 or 1400 MHz catalog flux densities. 

Density fluctuations in the ionized interstellar medium of our Galaxy can induce intensity fluctuations that may depress the flux density of a pulsar during an integration time. The characteristic time and frequency scale depends on many factors including the distance of the pulsar, the turbulent properties of the gas along the line of sight, and the relative velocities of the pulsar and the ionized gas \citep{cwf+91}. To estimate the magnitude of strong scattering on the phase-averaged pulsar flux densities we followed the method of \citet{kcac98}. We first estimated the scattering bandwidth and scattering time at 150 MHz for each pulsar using the NE2001 model of \citet{cl02}. Typical scintillation timescales and bandwidths at these frequencies are small, of order 1 minute and below 1 MHz, respectively. Our values are similar to the values estimated at the same frequency by \citet{bkk+15}. We then estimate the number of ``scintles'' that are averaged over the observed bandwidth and the duration of the observation. The intensity modulation is equal to the square root of this value. The observed bandwidth is given in \S\ref{survey} as 16.7 MHz. The duration of the GMRT observations are more difficult to estimate. The total integration on source is 15 minutes but it is split into 3--5 short snapshots spaced over a full night's observing. As an added complication, the image mosaics are additions of many overlapping fields and so it is possible that a single pixel may contain observations from more than one night. This sampling has the effect of smoothing out any large intensity modulations, so as a (pessimistic) estimate we take the duration as 15 min but we recognize that there may be additional temporal smoothing. Our results by and large suggest that the TGSS ADR pulsar flux densities are only being weakly modulated by scintillation in most cases. There are pulsars that are predicted to be undergoing strong diffractive scintillation at this frequency (e.g. PSR\,B0950+08 and PSR\,1929+10) and there are diffraction spikes centered on the MSP PSR\,B1937+12, likely caused by intensity variations on timescales comparable to the dump time. However, we can find no systematic trend for the non-detected pulsars to have greater predicted modulations from scattering.

Summarizing, we find that the bright cataloged pulsars with no TGSS ADR counterpart may be due to a combination of effects. There is evidence that the non-detections at 150 MHz have more shallow spectral indices than average, and that some of the non-detections are caused by high rms and confusion in the image plane. Strong intensity variations by interstellar scintillation is undoubtedly occurring for some pulsars but we cannot show that the non-detections differ from the detections in their scattering properties. The difficulty in estimating the true GMRT integration time for each pulsar may be masking this effect.

\subsection{Radio Quiet Pulsars}

\begin{figure}
\plotone{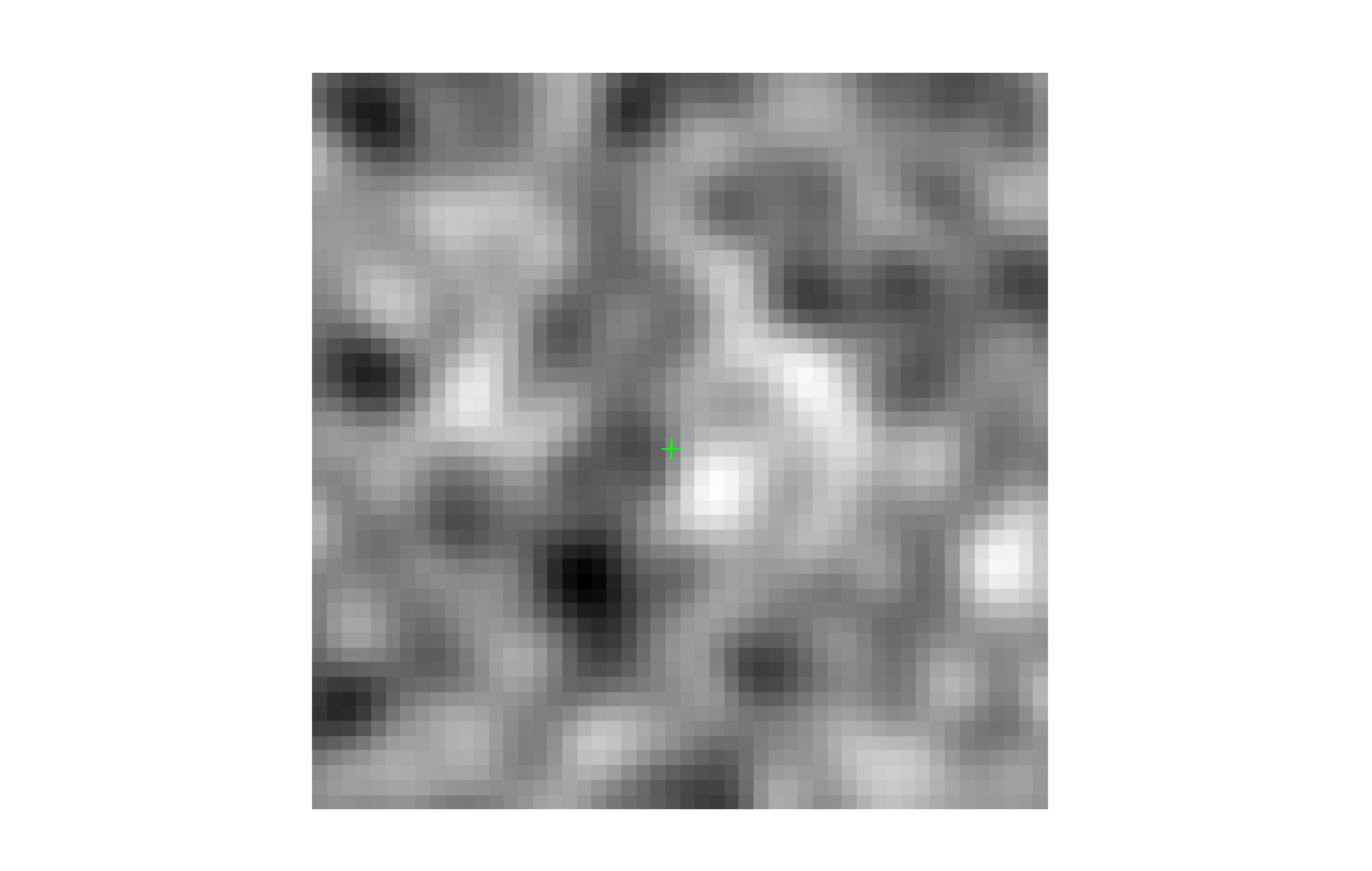}
\caption{A noise-weighted stack of all 30 gamma-ray radio-quiet pulsars at 150 MHz. The rms noise is \mjybeam{0.7} and the max/min is $\pm$\mjybeam{2.3}. The field of view is \amin{5}.
}\label{fig:stack}
\end{figure}

Our search did not find any significant radio emission at 150 MHz toward individual radio quiet pulsars, nor in a weighted stack of all 30 pulsars. The peak of the stacked image in Fig. \ref{fig:stack} is 0.1$\pm$0.7 mJy beam$^{-1}$, with upper limit (peak + 2$\sigma$) of $<$1.5 mJy beam$^{-1}$. Recall from \S\ref{quiet} that ``radio-quiet" pulsars are observationally defined as having a phase-averaged flux density at 1.4 GHz S$_\nu<$30 $\mu$Jy. The simplest hypothesis is that radio quiet pulsars are beamed away from the line of sight \citep{ckr+12}. 

Radio quiet gamma-ray pulsars, with Geminga as the prototype, are expected, given what we know about the structure of neutron star magnetospheres \citep{car14}. The radio emission is thought to originate further down the poles than the gamma-rays, and thus the radio will be beamed into a narrowing opening angle, increasing the probability that the beam sweeps out away from the observer's line of sight. However, it is well-known that both the radio pulse width and the component separation are frequency dependent \citep{thor91,mr02}. As noted by \citet{mad12}, this widening of radio beams at low frequencies might be used to detect radio quiet, gamma-ray pulsars. Such pulsars would be recognized in the image plane as having a spectral index that is much steeper than the canonical value. PSR\,B1943+10 may be thought of the prototype of such systems, bright at 400 MHz and below but weak at 1.4 GHz, with a  spectral index $\alpha$ steeper than $-3.0$ \citep[see Table \ref{tab:bright};][]{wcl+99}. The lower limit estimate on spectral index that we derive from the weighted stack at 150 MHz, assuming the defining radio-quiet 1.4 GHz flux density of 30 $\mu$Jy, gives $\alpha>-1.75\pm0.20$.  This is a spectral index limit that is well within the canonical value for normal pulsars. Thus we find no evidence that these gamma-ray pulsars have radio beams that sweep close to our line of sight.

%See this paper for gamma-rays and radio pulsars
% \bibitem[Ravi et al.(2010)]{2010ApJ...716L..85R} Ravi, V., Manchester, R.~N., \& Hobbs, G.\ 2010, \apjl, 716, L85 

\section{Conclusion}\label{sec:conclude}

We have identified nearly 300 pulsars at 150 MHz based in their phase-averaged emission on all-sky images. This imaging approach is complementary to pulsation studies since it is not affected by {pulse scatter} broadening or dispersion, making it sensitive to both normal and millisecond pulsars equally. Our sample includes many southern pulsars which are being detected at low radio frequencies for the first time. We anticipate that these 150 MHz flux densities will be used to study large numbers of pulsar over a wider frequency range than has hitherto been possible, and to addresses questions about the incidence and origins of low-frequency spectral turnovers. Accurate calibration between telescopes remains an important issue and we have identified a discrepancy between the flux densities of pulsars in common between GMRT and LOFAR. We suggest that the LOFAR sample may be overestimating the flux density scale by about a factor of two. It should be straightforward to test this hypothesis by observing a sample of pulsars with LOFAR in both imaging and phase-binning modes, calibrating the interferometric data in the standard way to allow proper comparison with each other and with the GMRT.

We have carried out a preliminary spectral index study of our sample. Generally there is good agreement with past work, except that we find a curious preponderance of gamma-ray MSPs with unusually steep spectral indices ($\alpha\leq-2.5$). Regardless of its origins, this suggests a possible way to identify new MSP candidates in \fermi unassociated sources on the basis of their unusually steep spectrum at low radio frequencies. Such pulsars may have been missed in radio pulsation searches due to propagation effects caused by the interstellar medium or they may be in binary systems and thus more difficult to discover. In such cases, imaging \fermi error regions with LOFAR and the GMRT could provide accurate enough positions to enable blind gamma-ray searches for pulsations.

\acknowledgments

This research has made use of data and/or software provided by the High Energy Astrophysics Science Archive Research Center (HEASARC), which is a service of the Astrophysics Science Division at NASA/GSFC and the High Energy Astrophysics Division of the Smithsonian Astrophysical Observatory. DAF thanks T. Readhead and S. Kulkarni  for their hospitality at Caltech while this work was being written up.
HTI acknowledges financial support through the NL-SKA roadmap project funded by the NWO. We thank A. Bilous and J. Hessels for sharing their knowledge of LOFAR pulsar flux density calibration.

{\it Facilities:} \facility{GMRT}, \facility{Fermi (LAT)}.

\clearpage

%\clearpage

\clearpage

%\begin{figure}
%\plotone{stack-eps-converted-to.pdf}
%\caption{ADD PROPER FIGURE AND INFORMATION}\label{fig:faint}
%\end{figure}

% Used psrcat-allDec2015.fits with non-zero spectral indces. Bin=0.25. Norm=area.

\clearpage

\clearpage

\clearpage
\begin{deluxetable}{lrrccll}
\tabletypesize{\scriptsize}
\rotate
\tablecaption{Bright Radio Pulsars Detected at 150 MHz\label{tab:bright}}
\tablewidth{0pt}
\tablehead{
\colhead{PSR} & \colhead{Period} & \colhead{DM} & \colhead{S$_t$} & \colhead{S$_p$} &
\colhead{$\alpha$} & \colhead{Notes}\\
\colhead{Name} & \colhead{(msec)} & \colhead{(pc cm$^{-3}$)} & \colhead{(mJy)} & \colhead{(mJy/ba)} &
\colhead{ } & \colhead{ }\\
}
\startdata
J0030+0451 & 4.87 & 4.3 & 44.9$\pm$4.5 & 34.9$\pm$2.8 & -1.93$\pm$0.15 & 2FGL\,\,J0030.4+0450\\
B0031-07 & 942.95 & 10.9 & 516.5$\pm$6.4 & 466.7$\pm$2.8 & -1.72$\pm$0.13 & \\
J0034-0534 & 1.88 & 13.8 & 265.3$\pm$4.4 & 250.8$\pm$2.6 & -2.72$\pm$0.08 & 2FGL\,J0034.4-0534, binary\\
B0114+58 & 101.44 & 49.4 & 71.3$\pm$5.4 & 55.1$\pm$3.4 & -2.45$\pm$0.04 & \\ 
B0136+57 & 272.45 & 73.8 & 154.5$\pm$4.9 & 146.3$\pm$2.9 & -1.57$\pm$0.06 & \\
B0138+59 & 1222.95 & 34.9 & 138.2$\pm$4.4 & 121.2$\pm$2.6 & -1.53$\pm$0.10 & \\
B0144+59 & 196.32 & 40.1 & 33.1$\pm$3.5 & 14.0$\pm$2.6 & -1.23$\pm$0.08 & \\
B0148-06 & 1464.66 & 25.7 & 28.9$\pm$5.0 & 26.1$\pm$3.0 & -1.39$\pm$0.27 & \\
B0149-16 & 832.74 & 11.9 & 88.4$\pm$7.0 & 81.8$\pm$4.1 & -1.83$\pm$0.12 & \\
J0218+4232 & 2.32 & 61.3 & 432.0$\pm$5.9 & 395.2$\pm$3.5 & -2.76$\pm$0.11 & 2FGL\,J0218.1+4233, binary\\
J0248+6021 & 217.09 & 370.0 & 152.3$\pm$5.4 & 131.6$\pm$3.2 & -1.08$\pm$0.10 & 2FGL\,J0248.1+6021\\
B0301+19 & 1387.58 & 15.7 & 30.3$\pm$4.5 & 22.5$\pm$2.9 & -1.04$\pm$0.15 & \\
B0320+39 & 3032.07 & 26.2 & 104.7$\pm$7.5 & 99.1$\pm$4.4 & -2.13$\pm$0.11 & \\
B0329+54 & 714.52 & 26.8 & 2156.9$\pm$11.2 & 1744.9$\pm$5.2 & -1.06$\pm$0.13 & \\
B0353+52 & 197.03 & 103.7 & 26.4$\pm$4.3 & 20.9$\pm$2.7 & -1.18$\pm$0.06 & \\
B0355+54 & 156.38 & 57.1 & 92.4$\pm$4.6 & 91.7$\pm$2.6 & -0.62$\pm$0.14 & \\
B0402+61 & 594.58 & 65.3 & 32.4$\pm$6.3 & 29.8$\pm$3.7 & -1.10$\pm$0.05 & \\
J0437-4715 & 5.76 & 2.6 & 313.9$\pm$8.9 & 311.9$\pm$4.4 & -0.33$\pm$0.11 & 2FGL\,J0437.3-4712, binary\\
B0447-12 & 438.01 & 37.0 & 64.1$\pm$4.2 & 61.4$\pm$2.4 & -1.91$\pm$0.06 & \\
B0450-18 & 548.94 & 39.9 & 111.2$\pm$6.4 & 98.9$\pm$3.8 & -1.36$\pm$0.08 & \\ 
B0450+55 & 340.73 & 14.6 & 31.2$\pm$4.5 & 29.8$\pm$2.6 & -0.39$\pm$0.11 & \\
J0459-2010 & 1133.08 & 21.0 & 62.3$\pm$7.2 & 54.4$\pm$4.3 & -2.16$\pm$0.04 & \\
B0458+46 & 638.57 & 42.2 & 25.0$\pm$4.4 & 19.5$\pm$2.7 & -1.03$\pm$0.05 & \\
B0523+11 & 354.44 & 79.3 & 31.8$\pm$5.6 & 29.6$\pm$3.3 & -1.34$\pm$0.05 & \\
B0531+21 & 33.39  & 56.8 & 151870$\pm$697 & 12510$\pm$40 & \nodata & Crab \\ % Added
B0559-05 & 395.97 & 80.5 & 32.5$\pm$5.5 & 31.1$\pm$3.2 & -1.15$\pm$0.05 & \\
J0613+3731 &  619.20  &  19.0 &  72.1$\pm$5.4 &  38.7$\pm$3.8 & \nodata & LOFAR PSR. Extended? \\ %added
J0621+1002 & 28.85 & 36.6 & 77.3$\pm$7.1 & 61.5$\pm$4.4 & -1.66$\pm$0.08 & \\
J0627+0706 & 475.88 & 138.1 & 56.4$\pm$7.5 & 44.8$\pm$4.6 & -1.33$\pm$0.13 & \\
B0628-28 & 1244.42 & 34.4 & 337.6$\pm$5.3 & 332.1$\pm$3.0 & -1.20$\pm$0.12 & \\
B0655+64 & 195.67 & 8.8 & 50.9$\pm$6.3 & 48.0$\pm$3.7 & -2.30$\pm$0.15 & \\
J0737-3039A/B & 22.7/2773.46 & 48.9 & 125.8$\pm$9.1 & 125.8$\pm$5.1 &  \nodata & Double PSR , Fermi PSR\\
%J0737-3039A & 22.7 & 48.9 & 125.8$\pm$9.1 & 125.8$\pm$5.1 & -1.95 & Double PSR\\
B0736-40 & 374.92 & 160.8 & 128.4$\pm$5.5 & 129.2$\pm$3.0 & -0.21$\pm$0.04 & \\
B0740-28 & 166.76 & 73.8 & 392.7$\pm$6.2 & 376.5$\pm$3.5 & -1.46$\pm$0.06 & 2FGL\,J0742.4-2821\\
B0808-47 & 547.2 & 228.3 & 291.3$\pm$9.5 & 270.3$\pm$4.9 & -2.05$\pm$0.04 & \\
B0809+74 & 1292.24 & 5.7 & 435.5$\pm$3.7 & 398.7$\pm$2.2 & -1.69$\pm$0.06 & \\
B0818-41 & 545.45 & 113.4 & 207.0$\pm$7.6 & 204.4$\pm$4.1 & -1.65$\pm$0.06 & \\
B0818-13 & 1238.13 & 40.9 & 147.3$\pm$5.5 & 145.7$\pm$3.1 & -1.36$\pm$0.08 & \\
B0820+02 & 864.87 & 23.7 & 39.7$\pm$4.1 & 35.3$\pm$2.4 & -1.47$\pm$0.21 & \\
B0823+26 & 530.66 & 19.5 & 227.3$\pm$5.8 & 214.2$\pm$3.4 & -1.40$\pm$0.10 & \\
B0833-45 & 89.33 & 68.0 & 10250.1$\pm$24.0 & 8358.8$\pm$8.7 & -1.00$\pm$0.04 & Vela SNR, 2FGL\,J0835.3-4510\\
B0834+06 & 1273.77 & 12.9 & 766.2$\pm$8.6 & 637.7$\pm$3.8 & -2.35$\pm$0.12 & \\
B0835-41 & 751.62 & 147.3 & 104.9$\pm$5.3 & 104.3$\pm$2.8 & -0.84$\pm$0.06 & \\
B0840-48 & 644.35 & 196.9 & 54.3$\pm$5.7 & 53.0$\pm$3.0 & -2.00$\pm$0.07 & \\
B0844-35 & 1116.1 & 94.2 & 18.3$\pm$5.0 & 16.6$\pm$2.9 & -0.86$\pm$0.06 & \\
B0853-33 & 1267.54 & 86.6 & 44.7$\pm$6.6 & 44.5$\pm$3.6 & -1.93$\pm$0.08 & \\
J0857-4424 & 326.77 & 184.4 & 106.0$\pm$6.1 & 32.4$\pm$4.8 & -2.14$\pm$0.07 & \\ % Why is S_t>>S_p????
B0905-51 & 253.56 & 103.7 & 118.3$\pm$13.4 & 118.2$\pm$6.2 & -1.14$\pm$0.06 & \\
B0906-17 & 401.63 & 15.9 & 46.0$\pm$7.5 & 39.0$\pm$4.5 & -1.19$\pm$0.10 & \\
B0919+06 & 430.63 & 27.3 & 216.3$\pm$7.6 & 188.5$\pm$4.6 & -1.76$\pm$0.10 & Fermi PSR\\
B0943+10 & 1097.71 & 15.3 & 127.9$\pm$22.2 & 123.9$\pm$15.6 & -3.44$\pm$0.12 & \\ % Added
B0950-38 & 1373.82 & 167.0 & 30.8$\pm$5.9 & 24.0$\pm$3.6 & -1.31$\pm$0.04 & \\
B0950+08 & 253.07 & 3.0 & 655.9$\pm$14.1 & 602.0$\pm$8.4 & -0.92$\pm$0.05 & \\
B0957-47 & 670.09 & 92.7 & 66.1$\pm$3.1 & 63.6$\pm$1.6 & -1.44$\pm$0.04 & \\ 
J1022+1001 & 16.45 & 10.3 & 33.8$\pm$5.3 & 31.1$\pm$3.1 & -0.77$\pm$0.39 & \\
J1034-3224 & 1150.59 & 50.8 & 33.7$\pm$6.2 & 24.2$\pm$3.8 & -0.88$\pm$0.04 & \\
J1045-4509 & 7.47 & 58.1 & 55.2$\pm$5.7 & 50.7$\pm$3.1 & -1.35$\pm$0.12 & \\
B1112+50   & 1656.44  &   9.2 &	197.6$\pm$3.7 &	169.6$\pm$2.2 &	 -1.87$\pm$0.29  & \\  %Added
B1114-41 & 943.16 & 40.5 & 196.9$\pm$5.3 & 190.9$\pm$2.8 & -1.87$\pm$0.04 & \\
B1133+16 & 1187.91 & 4.8 & 644.3$\pm$6.1 & 590.6$\pm$3.6 & -1.34$\pm$0.08 & \\
B1237+25 & 1382.45 & 9.3 & 136.1$\pm$5.8 & 127.8$\pm$3.4 & -1.17$\pm$0.10 & \\
B1257+12  &    6.22  & 10.2  & 136.8$\pm$5.9  & 127.9$\pm$3.3 & -1.89$\pm$0.22 &  \\ % Added
B1322+83 & 670.04 & 13.3 & 18.3$\pm$5.0 & 15.5$\pm$3.0 & -1.35$\pm$0.04 & \\
B1352-51 & 644.3 & 112.1 & 105.0$\pm$6.8 & 109.0$\pm$3.1 & -2.21$\pm$0.04 & \\
J1418-3921 & 1096.81 & 60.5 & 82.4$\pm$7.8 & 66.4$\pm$4.5 & -1.82$\pm$0.04 & \\
B1504-43 & 286.76 & 48.7 & 200.3$\pm$10.3 & 175.4$\pm$5.6 & -2.58$\pm$0.22 & \\
B1508+55 & 739.68 & 19.6 & 988.2$\pm$7.2 & 848.3$\pm$4.4 & -2.16$\pm$0.07 & \\
B1530+27 & 1124.84 & 14.7 & 39.7$\pm$3.5 & 32.7$\pm$2.2 & -1.75$\pm$0.17 & \\
B1534+12 & 37.9 & 11.6 & 32.8$\pm$5.1 & 29.0$\pm$3.1 & -1.79$\pm$0.15 & Relativistic binary\\
B1540-06 & 709.06 & 18.4 & 145.4$\pm$5.6 & 140.2$\pm$3.3 & -1.92$\pm$0.16 & \\
B1541+09 & 748.45 & 35.0 & 368.7$\pm$4.5 & 349.9$\pm$2.6 & -1.85$\pm$0.08 & \\
J1543-5149 & 2.06 & 50.9 & 157.8$\pm$12.2 & 151.5$\pm$5.8 & -2.53$\pm$0.22 & 3FGL J1542.9-5141, binary\\
B1556-44 & 257.06 & 56.1 & 44.5$\pm$9.4 & 46.7$\pm$4.7 & -0.05$\pm$0.04 & \\
B1557-50 & 192.6 & 260.6 & 190.6$\pm$13.5 & 172.4$\pm$7.0 & -1.08$\pm$0.06 & \\
B1600-27 & 778.31 & 46.2 & 106.5$\pm$7.6 & 105.4$\pm$4.3 & -1.85$\pm$0.09 & \\
B1600-49 & 327.42 & 140.8 & 82.3$\pm$15.3 & 87.4$\pm$7.0 & -1.21$\pm$0.06 & \\
B1604-00 & 421.82 & 10.7 & 198.1$\pm$11.1 & 190.4$\pm$6.5 & -1.65$\pm$0.18 & \\
B1609-47   &  383.38  & 161.2 &  64.4$\pm$10.7&  67.2$\pm$5.3 &  -1.68$\pm$0.06  & \\ %added
B1610-50  &  231.69  & 528.8 & 297.0$\pm$28.1 & 276.5$\pm$14.0 & -2.16$\pm$0.07 &  \\ % Added. 
B1612+07 & 1206.8 & 21.4 & 31.6$\pm$5.8 & 34.6$\pm$3.3 & -1.77$\pm$0.22 & \\
B1620-09 & 1276.45 & 68.2 & 37.2$\pm$4.3 & 27.3$\pm$2.7 & -1.85$\pm$0.08 & \\
B1629-50 & 352.14 & 398.4 & 156.9$\pm$15.0 & 158.0$\pm$7.4 & -1.48$\pm$0.06 & \\
B1633+24 & 490.51 & 24.3 & 48.3$\pm$4.7 & 42.2$\pm$2.8 & -2.15$\pm$0.04 & \\
J1643-1224 & 4.62 & 62.4 & 186.7$\pm$5.7 & 167.8$\pm$3.4 & -1.64$\pm$0.06 & \\
B1641-45 & 455.06 & 478.8 & 108.4$\pm$20.3 & 107.7$\pm$10.2 & 0.47$\pm$0.04 & \\
B1642-03 & 387.69 & 35.8 & 429.6$\pm$5.8 & 381.3$\pm$3.5 & -1.35$\pm$0.13 & \\
J1645+1012 & 410.86 & 36.3 & 44.8$\pm$5.1 & 35.8$\pm$3.2 & -2.82$\pm$0.05 & \\
B1648-42 & 844.08 & 482.0 & 2470.5$\pm$15.6 & 2393.4$\pm$8.2 & -2.26$\pm$0.06 & \\
B1700-32 & 1211.79 & 110.3 & 162.7$\pm$8.0 & 132.4$\pm$4.8 & -1.37$\pm$0.06 & \\
B1700-18 & 804.34 & 49.6 & 41.3$\pm$5.8 & 40.8$\pm$3.3 & -1.83$\pm$0.13 & \\
B1702-19 & 298.99 & 22.9 & 52.6$\pm$5.4 & 30.5$\pm$3.6 & -0.84$\pm$0.17 & Fermi PSR\\ % near a mosaic join
J1705-3423 & 255.43 & 146.4 & 33.1$\pm$10.2 & 40.0$\pm$5.3 & -0.93$\pm$0.06 & \\
B1703-40 & 581.02 & 360.0 & 956.5$\pm$22.4 & 725.3$\pm$13.6 & -2.19$\pm$0.06 & SNR G345.7-00.2\\
B1706-16 & 653.05 & 24.9 & 82.5$\pm$6.2 & 85.2$\pm$3.5 & -1.35$\pm$0.22 & \\
B1714-34 & 656.3 & 587.7 & 149.5$\pm$15.8 & 142.6$\pm$9.0 & -1.71$\pm$0.06 & \\
B1717-16 & 1565.6 & 44.8 & 49.1$\pm$6.3 & 42.7$\pm$3.8 & -1.70$\pm$0.16 & \\
B1717-29 & 620.45 & 42.6 & 342.5$\pm$9.9 & 313.2$\pm$5.7 & -2.28$\pm$0.06 & \\
B1718-32 & 477.16 & 126.1 & 383.4$\pm$13.0 & 363.6$\pm$7.4 & -2.12$\pm$0.07 & \\
B1718-02 & 477.72 & 67.0 & 56.6$\pm$12.4 & 55.9$\pm$5.6 & -1.81$\pm$0.10 & \\ %added
J1730-2304 & 8.12 & 9.6 & 84.1$\pm$10.2 & 75.1$\pm$6.0 & -1.37$\pm$0.22 & Fermi PSR\\
B1727-47 & 829.83 & 123.3 & 237.6$\pm$8.3 & 217.0$\pm$4.4 & -1.34$\pm$0.04 & \\
B1737+13 & 803.05 & 48.7 & 43.9$\pm$5.6 & 38.6$\pm$3.4 & -1.08$\pm$0.07 & \\
B1740-31 & 2414.58 & 193.1 & 83.6$\pm$14.2 & 88.0$\pm$7.7 & -1.69$\pm$0.06 & \\
B1740-13 & 405.34 & 116.3 & 28.9$\pm$7.4 & 29.1$\pm$4.2 & -1.82$\pm$0.10 & \\
J1747-4036 & 1.65 & 153.0 & 580.9$\pm$23.6 & 426.3$\pm$10.7 & -2.81$\pm$0.22 & 3FGL\,J1747.6-4037\\
B1747-46 & 742.35 & 20.4 & 326$\pm$43.1 & 290.1$\pm$28.0 & -1.56$\pm$0.04 & \\ %Added
B1749-28 & 562.56 & 50.4 & 2591.9$\pm$30.0 & 2402.5$\pm$12.7 & -2.23$\pm$0.06 & \\
J1754-3443 & 361.69 & 187.7 & 74.3$\pm$9.3 & 52.6$\pm$5.8 & -2.25$\pm$0.07 & \\
J1758+3030 & 947.26 & 34.9 & 29.4$\pm$4.3 & 25.1$\pm$2.6 & -1.22$\pm$0.09 & \\
B1756-22 & 460.97 & 177.2 & 138.6$\pm$17.0 & 140.3$\pm$9.6 & -2.09$\pm$0.06 & \\
B1757-24 & 124.92 & 289.0 & 82.0$\pm$24.0 & 86.8$\pm$13.4 & -2.05$\pm$0.06 & SNR G5.4-1.2,  Fermi PSR\\
B1758-29 & 1081.91 & 125.6 & 58.4$\pm$9.0 & 37.5$\pm$5.9 & -1.56$\pm$0.06 & \\
J1802-2124 & 12.65 & 149.6 & 71.5$\pm$13.6 & 56.1$\pm$8.4 & -2.03$\pm$0.07 & \\
B1804-08 & 163.73 & 112.4 & 94.5$\pm$12.4 & 66.1$\pm$7.9 & -0.82$\pm$0.06 & \\
J1809-3547 & 860.39 & 193.8 & 103.9$\pm$9.3 & 93.2$\pm$5.3 & -1.63$\pm$0.04 & \\
J1810+1744 & 1.66 & 39.7 & 299.4$\pm$6.0 & 260.1$\pm$3.6 & -2.76$\pm$0.22 & 2FGL 1810.7+1742\\
J1811-1925 & 64.67 & 0.0 & 62992.6$\pm$92.7 & 2256.0$\pm$14.5 & \nodata & SNR G11.2-0.3\\
B1813-17 & 782.31 & 525.5 & 35.2$\pm$13.8 & 45.2$\pm$7.2 & -1.51$\pm$0.06 & \\
J1816+4510 & 3.19 & 38.9 & 123.9$\pm$6.9 & 116.6$\pm$4.1 & -3.46$\pm$0.10 & 1FGL\,J1816.7+4509, eclipsing binary\\
B1818-04 & 598.08 & 84.4 & 974.9$\pm$8.4 & 921.2$\pm$5.0 & -2.27$\pm$0.06 & \\
B1821+05 & 752.91 & 66.8 & 168.8$\pm$8.4 & 150.3$\pm$5.0 & -2.06$\pm$0.11 & \\
%J1823-3021D & 3.02 & 86.8 & 384.0$\pm$12.7 & 297.4$\pm$5.5 & \nodata & GC NGC6624 \\
B1820-30A & 5.44 & 86.9 & 384.0$\pm$12.7 & 297.4$\pm$5.5 & \nodata & GC NGC6624, Fermi PSR\\
B1820-31 & 284.05 & 50.2 & 171.6$\pm$9.7 & 153.9$\pm$5.6 & -1.89$\pm$0.11 & \\
B1821-19 & 189.33 & 224.6 & 100.5$\pm$8.3 & 92.0$\pm$4.9 & -1.35$\pm$0.06 & \\
B1821-24A & 3.05 & 119.9 & 405.6$\pm$10.6 & 350.4$\pm$6.3 & -2.38$\pm$0.10 & GC M28, Fermi PSR\\
B1822-09 & 769.01 & 19.4 & 411.7$\pm$10.9 & 363.9$\pm$6.5 & -1.58$\pm$0.06 & \\
B1826-17 & 307.13 & 217.1 & 102.2$\pm$7.5 & 102.7$\pm$4.3 & -1.16$\pm$0.06 & \\
J1833-1034 & 61.88 & 169.5 & 7116.8$\pm$28.4 & 1865.4$\pm$10.8 & \nodata & SNR G21.5-0.9, 2FGL\,J1833.6-1032\\
B1831-03 & 686.7 & 234.5 & 229.6$\pm$11.1 & 218.8$\pm$6.5 & -1.97$\pm$0.06 & \\
B1831-04 & 290.11 & 79.3 & 487.3$\pm$11.3 & 446.2$\pm$6.7 & -2.05$\pm$0.06 & \\
B1832-06   &  305.83  &	472.9 &	189.8$\pm$19.3&	144.7$\pm$12.1&	 -2.23$\pm$0.06  &  \\ %added
%J1835-1106&  165.91  & 132.7 &  92.8$\pm$11.5  &  80.8$\pm$6.9  & -1.68$\pm$0.10 &  \\ % added 
J1835-1106 & 165.91 & 132.7 & 92.8$\pm$11.5 & 80.8$\pm$6.9 & -1.68$\pm$0.06 & Fermi PSR\\
B1834-10 & 562.71 & 317.0 & 65.4$\pm$15.0 & 55.3$\pm$9.0 & -1.29$\pm$0.06 & \\
B1839+56 & 1652.86 & 26.8 & 65.5$\pm$5.9 & 51.1$\pm$3.7 & -1.25$\pm$0.12 & \\
J1843-1113 & 1.85 & 60.0 & 139.9$\pm$8.8 & 123.1$\pm$5.3 & -3.24$\pm$0.10 & 3FGL J1843.6-1114\\
B1842+14 & 375.46 & 41.5 & 105.4$\pm$7.3 & 89.9$\pm$4.5 & -1.90$\pm$0.10 & \\
J1846-0749 & 350.11 & 388.3 & 78.2$\pm$8.6 & 64.8$\pm$5.2 & -2.42$\pm$0.04 & \\
B1844-04 & 597.77 & 142.0 & 945.1$\pm$14.2 & 892.8$\pm$8.4 & -2.41$\pm$0.06 & \\
B1845-01 & 659.43 & 159.5 & 420.0$\pm$17.9 & 393.8$\pm$10.6 & -1.74$\pm$0.06 & \\
B1846-06 & 1451.32 & 148.2 & 202.8$\pm$7.5 & 195.1$\pm$4.4 & -2.23$\pm$0.06 & \\
B1848+04 & 284.7 & 115.5 & 93.4$\pm$9.2 & 81.9$\pm$5.5 & -2.22$\pm$0.07 & \\
B1857-26 & 612.21 & 38.0 & 408.2$\pm$15.3 & 389.2$\pm$8.8 & -1.54$\pm$0.05 & \\
B1859+03 & 655.45 & 402.1 & 437.1$\pm$17.3 & 360.8$\pm$10.6 & -2.08$\pm$0.06 & \\
J1902-5105 & 1.74 & 36.2 & 317.2$\pm$6.6 & 246.3$\pm$3.8 & -2.50$\pm$0.19 & 3FGL\,J1902.0-5107, binary\\
B1900-06 & 431.89 & 195.6 & 149.9$\pm$7.2 & 130.4$\pm$4.3 & -1.91$\pm$0.04 & \\
B1902-01 & 643.18 & 229.1 & 37.9$\pm$7.3 & 34.1$\pm$4.4 & -1.66$\pm$0.06 & \\
B1905+39 & 1235.76 & 31.0 & 38.1$\pm$5.3 & 36.8$\pm$3.1 & -1.37$\pm$0.08 & \\
B1907+00 & 1016.95 & 112.8 & 59.0$\pm$8.1 & 40.6$\pm$5.3 & -1.89$\pm$0.07 & \\
B1907+10 & 283.64 & 150.0 & 325.6$\pm$11.6 & 298.6$\pm$6.9 & -2.30$\pm$0.06 & \\
B1907+03 & 2330.26 & 82.9 & 165.1$\pm$14.0 & 161.0$\pm$8.1 & -2.10$\pm$0.06 & \\
B1907-03 & 504.6 & 205.5 & 123.7$\pm$6.5 & 119.3$\pm$3.8 & -2.42$\pm$0.07 & \\
J1911-1114 & 3.63 & 31.0 & 62.4$\pm$7.9 & 50.9$\pm$4.8 & -2.16$\pm$0.18 & \\
B1911-04 & 825.94 & 89.4 & 527.7$\pm$8.8 & 474.1$\pm$3.6 & -2.14$\pm$0.07 & \\
J1916+0844 & 440.0 & 339.4 & 42.3$\pm$8.6 & 38.3$\pm$5.1 & -2.04$\pm$0.07 & \\
B1914+09 & 270.25 & 61.0 & 63.8$\pm$10.1 & 52.5$\pm$6.2 & -1.90$\pm$0.06 & \\
B1915+13 & 194.63 & 94.5 & 149.6$\pm$21.3 & 107.7$\pm$13.5 & -1.95$\pm$0.06 & \\
B1918+19 & 821.04 & 153.9 & 71.2$\pm$6.7 & 66.0$\pm$3.9 & -1.67$\pm$0.09 & \\
B1919+21 & 1337.3 & 12.4 & 1169.1$\pm$11.3 & 882.0$\pm$4.3 & -2.36$\pm$0.04 & ISS\\
B1920+21 & 1077.92 & 217.1 & 131.3$\pm$9.2 & 98.4$\pm$3.8 & \nodata & Extended Em.\\
J1928+1923 & 817.33 & 476.0 & 32.1$\pm$5.8 & 32.9$\pm$3.3 & -1.75$\pm$0.04 & \\
J1930+1852 & 136.86 & 308.0 & 463.9$\pm$7.5 & 79.5$\pm$4.2 & \nodata & SNR G54.1+0.3\\
B1929+20 & 268.22 & 211.2 & 117.2$\pm$6.6 & 106.8$\pm$3.9 & -2.05$\pm$0.15 & \\
B1929+10 & 226.52 & 3.2 & 258.4$\pm$8.2 & 251.2$\pm$4.8 & -0.88$\pm$0.06 & \\
B1933+16 & 358.74 & 158.5 & 159.2$\pm$10.3 & 128.9$\pm$6.4 & -0.60$\pm$0.08 & \\
B1937+21 & 1.56 & 71.0 & 3619.8$\pm$19.2 & 2757.2$\pm$5.5 & -2.51$\pm$0.17 & 1FGL 1938.2+2125c, ISS\\
B1940-12 & 972.43 & 28.9 & 36.8$\pm$5.9 & 35.7$\pm$3.4 & -1.77$\pm$0.13 & \\
J1944+0907 & 5.19 & 24.3 & 95.3$\pm$8.1 & 80.5$\pm$4.9 & -3.03$\pm$0.06 & \\
B1946+35 & 717.31 & 129.1 & 462.3$\pm$10.6 & 422.8$\pm$6.3 & -1.80$\pm$0.06 & \\
B1951+32 & 39.53 & 45.0 & 873.2$\pm$37.9 & 370.4$\pm$16.6 & \nodata &  CTB\,80 PWN\\% added
B1953+29 & 6.13 & 104.5 & 56.7$\pm$5.1 & 48.1$\pm$3.1 & -1.77$\pm$0.09 & \\
B1957+20 & 1.61 & 29.1 & 398.1$\pm$7.0 & 381.2$\pm$4.1 & -3.09$\pm$0.22 & 2FGL\,J1959.5+2047, eclipsing binary\\
J2010+2845 & 565.37 & 112.5 & 45.4$\pm$4.4 & 25.8$\pm$3.0 & -2.09$\pm$0.05 & ISS\\
B2016+28 & 557.95 & 14.2 & 282.0$\pm$10.2 & 206.6$\pm$4.7 & -1.00$\pm$0.19 & \\
B2020+28 & 343.4 & 24.6 & 72.0$\pm$7.4 & 66.6$\pm$4.4 & -0.29$\pm$0.04 & \\
B2027+37 & 1216.8 & 190.7 & 132.0$\pm$14.1 & 105.5$\pm$8.7 & -2.41$\pm$0.08 & \\
J2043+2740 & 96.13 & 21.0 & 98.2$\pm$6.4 & 67.2$\pm$4.2 & -1.92$\pm$0.04 & 2FGL\,J2043.7+2743\\
B2043-04 & 1546.94 & 35.8 & 26.2$\pm$4.7 & 23.3$\pm$2.8 & -1.23$\pm$0.13 & \\
B2045-16 & 1961.57 & 11.5 & 360.1$\pm$9.9 & 291.9$\pm$4.5 & -1.49$\pm$0.14 & \\
J2051-0827 & 4.51 & 20.7 & 22.6$\pm$4.7 & 19.5$\pm$2.8 & -0.93$\pm$0.10 & Fermi PSR, binary\\
B2053+36 & 221.51 & 97.3 & 64.4$\pm$6.7 & 57.3$\pm$4.0 & -1.44$\pm$0.05 & \\
B2106+44 & 414.87 & 139.8 & 76.2$\pm$7.7 & 61.3$\pm$4.8 & -1.19$\pm$0.05 & \\
B2111+46 & 1014.68 & 141.3 & 461.7$\pm$7.3 & 442.0$\pm$4.3 & -1.43$\pm$0.05 & \\
J2139+2242 & 1083.51 & 44.1 & 26.8$\pm$4.8 & 24.3$\pm$2.8 & \nodata & \\
J2145-0750 & 16.05 & 9.0 & 237.4$\pm$5.4 & 220.2$\pm$3.2 & -1.47$\pm$0.61 & \\
B2148+63 & 380.14 & 128.0 & 32.7$\pm$5.2 & 27.7$\pm$3.2 & -1.08$\pm$0.07 & \\
B2152-31 & 1030.0 & 14.9 & 32.0$\pm$6.2 & 38.7$\pm$3.2 & -1.86$\pm$0.26 & \\
B2154+40 & 1525.27 & 70.9 & 87.3$\pm$9.0 & 73.2$\pm$5.5 & -0.73$\pm$0.07 & \\
J2215+5135 & 2.61 & 69.2 & 87.5$\pm$6.0 & 89.5$\pm$3.4 & -2.92$\pm$0.22 & 2FGL J2215.7+5135, eclipsing binary\\
B2217+47 & 538.47 & 43.5 & 592.4$\pm$6.6 & 578.9$\pm$3.8 & -2.37$\pm$0.04 & \\
B2224+65 & 682.54 & 36.1 & 62.1$\pm$4.6 & 55.7$\pm$2.8 & -1.54$\pm$0.10 & Guitar Nebula\\
J2229+6114 & 51.62 & 205.0 & 77.3$\pm$6.5 & 68.2$\pm$3.9 & \nodata & SNR G106.6+2.9, 2FGL\,J2229.0+6114\\
B2227+61 & 443.05 & 124.6 & 53.4$\pm$6.4 & 46.3$\pm$3.9 & -1.88$\pm$0.07 & \\
J2234+2114 & 1358.75 & 35.1 & 13.9$\pm$4.0 & 13.4$\pm$2.3 & -1.71$\pm$0.04 & \\
B2303+30 & 1575.89 & 49.6 & 41.9$\pm$3.6 & 32.3$\pm$2.3 & -1.32$\pm$0.11 & \\
B2306+55 & 475.07 & 46.5 & 72.6$\pm$14.5 & 69.9$\pm$10.0 & -1.63$\pm$0.08 & \\%added
B2310+42 & 349.43 & 17.3 & 55.2$\pm$6.5 & 53.3$\pm$3.8 & -0.58$\pm$0.12 & \\
J2317+1439 & 3.45 & 21.9 & 27.5$\pm$4.8 & 23.4$\pm$2.9 & -0.86$\pm$0.12 & Fermi PSR\\
J2319+6411 & 216.02 & 246.0 & 103.4$\pm$7.3 & 97.6$\pm$4.3 & -2.58$\pm$0.12 & \\
B2327-20 & 1643.62 & 8.5 & 197.7$\pm$6.2 & 173.4$\pm$3.7 & -1.88$\pm$0.15 & \\
B2334+61   &  495.37  &	 58.4 &	 80.1$\pm$10.2&	 62.3$\pm$6.3 &	 -1.81$\pm$0.10  & \\ % added
\enddata
%% Text for table notes should follow after the \enddata but before
%% the \end{deluxe table}. Make sure there is at least one \tablenotemark
%% in the table for each \tablenotetext.
\tablecomments{Columns from left to right include the common pulsar name, period, dispersion measure, total flux from TGSS ADR, peak flux density from TGSS ADR, spectral index, and notes on individual sources. Errors for the total flux density and peak flux are measured errors only. To get a conservative estimate of the uncertainty of the flux density scale, add a 10\% error in quadrature with these measured errors.}
\end{deluxetable}
\clearpage

\clearpage
\begin{deluxetable}{lrccc}
\tabletypesize{\scriptsize}
%\rotate
\tablecaption{Faint Radio Pulsars Detected at 150 MHz\label{tab:faint}}
\tablewidth{0pt}
\tablehead{
\colhead{PSR} & \colhead{Period} & \colhead{DM} & \colhead{S$_p$} &
\colhead{$\sigma$} \\
\colhead{Name} & \colhead{(msec)} & \colhead{(pc cm$^{-3}$)} & \colhead{(mJy/ba)} & \colhead{(mJy/ba)} 
}
\startdata
   B0011+47 & 1240.7 & 30.9 & 8.8 & 2.8\\
   B0045+33 & 1217.09 & 39.9 & 7.5 & 2.4\\
   B0059+65 & 1679.16 & 65.9 & 11.8 & 3.8\\
   B0105+65 & 1283.66 & 30.5 & 10.6 & 4.1\\
   J0134-2937 & 136.96 & 21.8 & 10.6 & 3.9\\
   J0337+1715 & 2.73 & 21.3 & 15.9 & 3.4\\
   B0339+53 & 1934.48 & 67.3 & 7.4 & 2.7\\
%   J0514-4002A & 4.99 & 52.1 & 7.6 & 2.9\\ % GC
   J0520-2553 & 241.64 & 33.8 & 8.5 & 3.0\\
   J0540+3207 & 524.27 & 62.0 & 12.0 & 3.6\\
   B0609+37 & 297.98 & 27.1 & 10.4 & 3.4\\
   J0646+0905 & 903.91 & 149.0 & 11.6 & 2.3\\
   B0656+14 & 384.89 & 14.0 & 5.7 & 2.2\\
   J0820-3921 & 1073.57 & 179.4 & 12.6 & 4.7\\
   J0820-3826 & 124.84 & 195.6 & 9.5 & 3.4\\
   B0906-49 & 106.75 & 180.4 & 16.9 & 4.0\\
   J1012+5307 & 5.26 & 9.0 & 6.0 & 2.3\\
   B1010-23 & 2517.95 & 22.5 & 17.8 & 6.3\\
   J1023+0038 & 1.69 & 14.3 & 9.4 & 2.9\\
   J1024-0719 & 5.16 & 6.5 & 8.3 & 2.5\\
   J1123-4844 & 244.84 & 92.9 & 4.5 & 1.5\\
   B1325-49 & 1478.72 & 118.0 & 8.2 & 2.9\\
   J1346-4918 & 299.63 & 74.4 & 7.8 & 2.5\\
   J1455-3330 & 7.99 & 13.6 & 21.3 & 6.7\\
   J1518+4904 & 40.93 & 11.6 & 9.2 & 3.6\\
   J1536-3602 & 1319.76 & 96.0 & 20.1 & 6.2\\
   J1555-0515 & 975.41 & 23.5 & 10.1 & 3.2\\
   J1557-4258 & 329.19 & 144.5 & 20.6 & 4.9\\
   J1612-2408 & 923.83 & 49.0 & 12.2 & 4.0\\
   J1614-3937 & 407.29 & 152.4 & 16.3 & 6.1\\
   J1627+1419 & 490.86 & 33.8 & 13.3 & 3.9\\
   J1630-4719 & 559.07 & 489.6 & 69.1 & 22.7\\
   J1637-4450 & 252.87 & 470.7 & 36.9 & 11.9\\
   B1635-45 & 529.12 & 258.9 & 39.5 & 14.2\\
   J1649+2533 & 1015.26 & 35.5 & 7.0 & 2.3\\
   J1653-2054 & 4.13 & 56.6 & 15.7 & 3.4\\
   J1655-3048 & 542.94 & 154.3 & 12.0 & 4.6\\
%   J1701-3006E & 3.23 & 113.8 & 17.8 & 6.5\\ %GC
 %  J1701-3006D & 3.42 & 114.2 & 17.8 & 6.5\\ %GC
   J1703-4851 & 1396.4 & 150.3 & 18.5 & 7.2\\
   J1709+2313 & 4.63 & 25.3 & 11.3 & 3.5\\
   J1716-4005 & 311.81 & 435.0 & 41.9 & 13.2\\
   B1727-33 & 139.46 & 259.0 & 43.8 & 13.2\\
   J1732-4156 & 323.43 & 228.7 & 20.1 & 7.3\\
   J1737-0811 & 4.18 & 55.3 & 21.5 & 7.8\\
   J1741+1351 & 3.75 & 24.2 & 9.2 & 2.7\\
   J1744-1134 & 4.07 & 3.1 & 10.2 & 3.7\\
   B1742-30 & 367.43 & 88.4 & 28.8 & 9.4\\
%   B1745-20A & 288.6 & 219.4 & 13.7 & 5.3\\ %GC
%   J1748-2021E & 16.26 & 224.1 & 14.0 & 5.3\\ %GC
%   J1748-2021B & 16.76 & 220.9 & 13.7 & 5.3\\ %GC
 %  J1750-3703D & 5.14 & 230.1 & 27.1 & 10.6\\ %GC
 %  J1750-3703C & 26.57 & 230.7 & 27.1 & 10.6\\ %GC
   J1750-3503 & 684.01 & 189.4 & 37.0 & 8.7\\
   J1752+2359 & 409.05 & 36.0 & 10.0 & 3.8\\
   B1754-24 & 234.1 & 179.5 & 37.1 & 12.7\\
   B1804-27 & 827.78 & 313.0 & 18.2 & 6.1\\
%   J1808-2024 & 7555.92 & 0.0 & 57.2 & 11.8\\ % is this SGR? Near cluster of radio sources.
%   J1811-1736 & 104.18 & 476.0 & 25.1 & 8.5\\  %local peak in extended emission
   B1817-18 & 309.9 & 436.0 & 19.0 & 7.2\\
   J1822-4209 & 456.51 & 72.5 & 13.1 & 5.1\\
   B1820-14 & 214.77 & 651.1 & 66.6 & 13.7\\
   J1832-0836 & 2.72 & 28.2 & 34.8 & 13.3\\
   B1829-10 & 330.35 & 475.7 & 60.3 & 17.0\\
   B1830-08 & 85.28 & 411.0 & 43.2 & 16.4\\
%   B1831-00 & 520.95 & 88.7 & 20.9 & 4.5\\ % src nearby but not likely PSR. NVSS src. Not in Kaplan et al.
   J1834-1710 & 358.31 & 123.8 & 12.4 & 3.6\\
   B1839+09 & 381.32 & 49.1 & 22.5 & 5.8\\
   J1843-0702 & 191.61 & 228.1 & 13.6 & 5.2\\
   J1843-0000 & 880.33 & 101.5 & 21.3 & 6.8\\
   B1841-05 & 255.7 & 412.8 & 40.3 & 12.5\\
   J1845-0743 & 104.69 & 281.0 & 11.3 & 4.1\\
   J1849+0409 & 761.19 & 56.1 & 13.7 & 4.8\\
   J1852-0127 & 428.98 & 431.0 & 30.3 & 8.2\\
   J1852+0008 & 467.89 & 254.9 & 34.0 & 10.6\\
   J1853+0505 & 905.14 & 279.0 & 23.2 & 6.6\\
   B1855+02 & 415.82 & 506.8 & 39.1 & 11.9\\
%   J1900-0051 & 385.19 & 136.8 & 30.5 & 12.2\\ %sidelobe of bright nearby src
   B1859+01 & 288.22 & 105.4 & 32.6 & 10.6\\
   B1900+06 & 673.5 & 502.9 & 28.5 & 9.7\\
%   J1904+0004 & 139.52 & 233.6 & 33.3 & 11.4\\ %local peak in extended emission
   J1908+0909 & 336.55 & 467.5 & 22.1 & 8.1\\
   B1913+10 & 404.55 & 241.7 & 23.9 & 6.0\\
   B1917+00 & 1272.26 & 90.3 & 24.7 & 6.3\\
   J1930-1852 & 185.52 & 42.9 & 8.9 & 3.5\\
   B1942-00 & 1045.63 & 59.7 & 12.5 & 5.0\\
   B1944+17 & 440.62 & 16.2 & 17.5 & 4.9\\
   B2000+32 & 696.76 & 142.2 & 10.2 & 3.4\\
   B2002+31 & 2111.26 & 234.8 & 10.8 & 3.1\\
   B2003-08 & 580.87 & 32.4 & 6.8 & 2.6\\
   J2013-0649 & 580.19 & 63.4 & 6.0 & 2.2\\
   B2021+51 & 529.2 & 22.6 & 16.9 & 6.2\\
   B2045+56 & 476.73 & 101.8 & 12.8 & 4.4\\
   B2053+21 & 815.18 & 36.4 & 11.0 & 3.7\\
   B2110+27 & 1202.85 & 25.1 & 11.1 & 3.1\\
   B2113+14 & 440.15 & 56.1 & 8.6 & 2.5\\
   J2205+1444 & 938.01 & 36.7 & 9.4 & 3.1\\
   J2215+1538 & 374.2 & 29.3 & 8.5 & 3.3\\
   J2229+2643 & 2.98 & 23.0 & 8.6 & 3.3\\
   B2315+21 & 1444.65 & 20.9 & 10.0 & 2.6\\
   B2323+63 & 1436.31 & 197.4 & 14.5 & 4.7\\
\enddata
%% Text for table notes should follow after the \enddata but before
%% the \end{deluxe table}. Make sure there is at least one \tablenotemark
%% in the table for each \tablenotetext.
\tablecomments{Columns from left to right include the common pulsar name, period, dispersion measure, peak flux density from TGSS ADR, and the local rms noise measured from the TGSS ADR image in the vicinity of each pulsar. Errors for the peak flux are measured errors only. To get a conservative estimate of the uncertainty of the flux density scale, add a 10\% error in quadrature with these measured errors.}
%\tablecomments{Table is published in its entirety in the electronic edition of the {\it Astrophysical Journal}.  A portion is  shown here for guidance regarding its form and content.}
%\tablenotetext{a}{Sample footnote for table that was generated with the deluxe table environment}
%\tablenotetext{b}{Another sample footnote for table.}
%\tablerefs{(a) \citep{naa+12}}
\end{deluxetable}
\clearpage

\clearpage
\begin{deluxetable}{ccccccc}
\tabletypesize{\scriptsize}
\rotate
\tablecaption{Radio Quiet Pulsars\label{tab:unknown}}
\tablewidth{0pt}
\tablehead{Name & RA (J2000) & Dec (J2000) & P & $\dot{P}$ & $\dot{E}$ & S$_p\pm\sigma$\\
            & (${\rm h}~{\rm m}~ {\rm s}$) & (${\circ} ~ {'} ~ {''}$) & (ms) & (s s$^{-1}$) & (erg s$^{-1}$) & (mJy beam$^{-1}$)
           }
\startdata
% J0002+6216 No accurate positions found.
J0007+7303  & 00:07:01.56 (0.01) &  +73:03:08.1 (0.1) & 315.89 &  3.57e-13 &  4.48e+35 &    $-0.8\pm2.2$\\  %kerr
J0106+4855  & 01:06:25:03 (0.05) &  +48:55:52.0 (0.6) &  83.16 &  4.28e-16 &  2.94e+34 &    $1.7\pm3.4$    \\ % Kerr
J0357+3205  & 03:57:52.33 (0.01) & +32:05:20.7 (0.3) & 444.10 &  1.31e-14 &  5.90e+33 &    $-2.5\pm2.7$  \\ %Kerr
%J0359+5414 No accurate positions found.
%J0554+3107 No accurate positions found.
J0622+3749  & 06:22:10:41 (0.07) & +37:49:14.6 (3.5) & 333.21 &  2.54e-14 &  2.71e+34 &  $3.8\pm3.3$ \\ %Kerr.Posn.Error.large.
%J0631+0646 No accurate positions found.
J0633+0632  & 06:33:44.14 (0.02) & +06:32:30.4 (0.3) & 297.40 &  7.96e-14 &  1.19e+35 &    $-4.6\pm4.4$ \\ %Kerr
J0633+1746  & 06:33:54.310 (0.001) & +17:46:14.60 (0.04) & 237.10 &  1.10e-14 &  3.25e+34 &    $2.7\pm2.9$ \\ %SLAC+HST
J0734$-$1559  & 07:34:45.70 (0.01) & -15:59:19.8 (0.3) & 155.14 &  1.25e-14 &  1.32e+35 &    $1.4\pm3.3$ \\ %Marelli
%J1623-5005 No accurate positions found.
%J1624-4041 No accurate positions found.
%J1650-4601 No accurate positions found.
J1620$-$4927  & 16:20:41.59 (0.09) &  -49:27:36.2 (1.7) & 171.93 &  1.05e-14 &  8.15e+34 &    $18\pm45$ \\ %SLAC
J1732$-$3131  & 17:32:33.55 (0.03) & -31:31:23.9 (0.5) & 196.54 &  2.80e-14 &  1.46e+35 &    $<-7.9\pm7.9$ \\ %Kerr.Chandra
J1741$-$2054   &  17:41:57.28 (0.02) &  -20:54:11.8 (0.3) & 413.70 & 1.70e-14 & 9.5e33 & $5.2\pm4.8$\\%Kerr.Chandra
%J1746$-$3239  & 17:46:55.14 (0.14) &  -32:39:36.4 (9.7) & 199.54 &  6.56e-15 &  3.26e+34 &    $\nodata$ \\ %SLACPosnErr.
J1803$-$2149  & 18:03:9.63 (0.01) & -21:49:00.9 (3.7) & 106.33 &  1.95e-14 &  6.41e+35 &    $12\pm10$ \\ %SLACPosnErr.
J1809$-$2332  & 18:09:50.25 (0.03) & -23:32:22.7 (0.1) & 146.79 &  3.44e-14 &  4.30e+35 &    $<9.9\pm8.4$ \\ %Kerr.Chandra.
J1813$-$1246  &  18:13:23.77 (0.01) & -12:46:00.6 (0.4) &  48.07 &  1.76e-14 &  6.24e+36 &    $5.2\pm7.7$ \\  %SLAC
J1826$-$1256  & 18:26:08.54 (0.01) &  -12:56:34.6 (0.1) & 110.23 &  1.21e-13 &  3.58e+36 &    $17.7\pm7.9$ \\ %Kerr.Chandra.
%J1827-1446 No accurate positions found.
J1836+5925  & 18:36:13.72 (0.02) & +59:25:30.1 (0.1) & 173.26 &  1.50e-15 &  1.14e+34 &    $2.9\pm3.6$ \\ %Kerr.Chandra
J1838$-$0537  & 18:38:56.2 (0.18) & -05:37:04.5 (2.7) & 145.71 &  4.65e-13 &  5.93e+36 &    $0.5\pm9.1$ \\    %1501.06125.Chandra
% J1844-0346 No accurate positions found.
J1846+0919  & 18:46:25.88 (0.03) & +09:19:49.8 (0.5) & 225.55 &  9.93e-15 &  3.42e+34 &    $3.8\pm6.1$ \\ %SLAC
%J1906+0722 No accurate positions found.
J1907+0602  & 19:07:54.76 (0.05) & +06:02:14.6 (0.7) & 106.64 &  8.67e-14 &  2.82e+36 &    $-0.7\pm8.1$ \\ %Kerr.Chandra
J1932+1916  &  19:32:19.78 (0.06) & +19:16:38.1 (1.6) & 208.21 & 9.32e-14 & 4.1e35 & $2.4\pm5.3$\\%SLAC
J1954+2836  &  19:54:19.14 (0.01) & +28:36:04.8 (0.1) &  92.71 &  2.12e-14 &  1.05e+36 &    $-2.5\pm2.8$ \\  %SLAC
J1957+5033  & 19:57:38.39 (0.07) & +50:33:21.2 (0.7) & 374.81 &  6.83e-15 &  5.12e+33 &    $1.0\pm3.7$ \\  %SLAC
J1958+2846  &  19:58:40.01 (0.02) & +28:45:55.1 (0.3) & 290.40 &  2.12e-13 &  3.42e+35 &    $1.4\pm3.2$ \\  %Kerr.Chandra
%J2017+3625 No accurate positions found.
J2021+4026  & 20:21:30.73 (0.02) &  +40:26:46.0 (0.3) & 265.32 &  5.42e-14 &  1.14e+35 &    $-32\pm39$ \\ %Kerr.Chandra
%J2022+3842 No accurate positions found.
J2028+3332  & 20:28:19.88 (0.01) & +33:32:04.2 (0.1) & 176.71 &  4.86e-15 &  3.48e+34 &    $-4.6\pm3.8$ \\ %SLAC
J2030+4415  & 20:30:51.40 (0.02) & +44:15:38.7 (0.3) & 227.07 &  6.49e-15 &  2.19e+34 &    $2.2\pm6.7$ \\ %SLAC
J2032+4127 & 20:32:13.14 (0.02) & +41:27:24.5 (0.3) & 143.25 & 1.25e-14 & 1.7e35 & $-11\pm7$\\%Kerr.Chandra
J2055+2539  & 20:55:48.95 (0.03) &  +25:39:58.9 (0.6) & 319.56 &  4.11e-15 &  4.97e+33 &    $2.3\pm5.8$\\ %SLAC
J2111+4606  & 21:11:24.13 (0.03) & +46:06:30.7 (0.7) & 157.83 &  1.43e-13 &  1.44e+36 &    $4.5\pm4.3$ \\ %SLAC
J2139+4716  & 21:39:55.89 (0.06) & +47:16:13.1 (0.7) & 282.85 &  1.80e-15 &  3.15e+33 &    $2.9\pm3.4$ \\ %SLAC
J2238+5903  & 22:38:28.90 (0.19) & +59:03:43.4 (1.5) & 162.74 &  9.70e-14 &  8.88e+35 &    $-4.8\pm3.4$ \\ %SLAC
\enddata
\tablenotetext{a}{Pulsar positions are taken from \citet{krj+15} and \citet{mmd+15} while the remaining pulsar parameters are from the ATNF Pulsar Catalog \citep{mhth05}.}
\end{deluxetable}

\end{document}